\documentclass[a4paper,fleqn,usenatbib]{mnras}

\usepackage{savesym}
\usepackage{amsmath}
\savesymbol{iint}
\usepackage{txfonts}
\restoresymbol{TXF}{iint}

\usepackage{amssymb,color,upgreek}

\usepackage{ulem}

\usepackage[T1]{fontenc}
\usepackage{ae,aecompl}

\usepackage{graphicx}	

\title[A dearth of small particles around WD 1145+017]{A dearth of small particles in the transiting material around the white dwarf WD 1145+017}

\author[Xu, Rappaport, \& van Lieshout et al.]{
S. Xu$^{1}$\thanks{E-mail: \href{sxu@gemini.edu}{sxu@gemini.edu}}, 
S. Rappaport$^2$, 
R. van Lieshout$^3$, 
A. Vanderburg$^4$, 
B. Gary$^5$, 
\newauthor N. Hallakoun$^{6,1}$, 
V. D. Ivanov$^{1,7}$,
M. C. Wyatt$^{3}$,
J. DeVore$^{8}$,
D. Bayliss$^{9}$, 
J. Bento$^{10}$, 
\newauthor A. Bieryla$^4$,
A. Cameron$^{11}$
J. M. Cann$^{12}$,  
B. Croll$^{13}$,
K. A. Collins$^{4}$, 
P. A. Dalba$^{13}$,
\newauthor J. Debes$^{14}$,
D. Doyle$^{15}$,
P. Dufour$^{16}$,
J. Ely$^{17}$,
N. Espinoza$^{18}$,
M. D. Joner$^{19}$ ,
\newauthor M. Jura$^{20}$,
T. Kaye$^{21}$, 
J. L. McClain$^{15,22}$,
P. Muirhead$^{13}$,
E. Palle$^{23,24}$,
P. A. Panka$^{12}$, 
\newauthor J. Provencal$^{25}$,
S. Randall$^{1}$,
J. E. Rodriguez$^{4}$, 
J. Scarborough$^{15}$,
R. Sefako$^{26}$,
\newauthor A. Shporer$^{27}$,
W. Strickland$^{15}$,
G. Zhou$^{4}$, 
B. Zuckerman$^{20}$
\newauthor \small\it{Affiliations are listed at the end of the paper.}
}

\date{Accepted 2017 November 14. Received 2017 November 14; in original form 2017 August 09}

\pubyear{2017}

\begin{document}
\label{firstpage}
\pagerange{\pageref{firstpage}--\pageref{lastpage}}
\maketitle

\newpage
\begin{abstract}
White dwarf WD~1145+017 is orbited by several clouds of dust, possibly emanating from actively disintegrating bodies. These dust clouds reveal themselves through deep, broad, and evolving transits in the star's light curve. Here, we report two epochs of multi-wavelength photometric observations of WD~1145+017, including several filters in the optical, K$_\mathrm{s}$ and 4.5~$\mu$m bands in 2016 and 2017. The observed transit depths are different at these wavelengths. However, after correcting for excess dust emission at K$_\mathrm{s}$ and 4.5~$\mu$m, we find the transit depths for the white dwarf itself are the same at all wavelengths, at least to within the observational uncertainties of $\sim$5\%-10\%. From this surprising result, and under the assumption of low optical depth dust clouds, we conclude that there is a deficit of small particles (with radii $s \lesssim$ 1.5~$\mu$m) in the transiting material. We propose a model wherein only large particles can survive the high equilibrium temperature environment corresponding to 4.5~hr orbital periods around WD~1145+017, while small particles sublimate rapidly. In addition, we evaluate dust models that are permitted by our measurements of infrared emission. 
\end{abstract}

\begin{keywords}
eclipses -- minor planets, asteroids: general -- stars: individual: WD~1145+017 -- white dwarfs.
\end{keywords}



\section{Introduction}
Recent studies show that relic planetary systems or their debris are widespread around white dwarfs. About 25-50\% of white dwarfs show ``pollution" from elements heavier than helium in their atmospheres \citep{Zuckerman2003, Zuckerman2010, Koester2014a}. The most heavily polluted white dwarfs often display excess infrared emission from a dust disc within the white dwarf's tidal radius \citep[e.g.][]{vonHippel2007, Farihi2009}. About 20\% of these dusty white dwarfs also display double-peaked calcium infrared triplet emission lines from orbiting gas debris which spatially coincides with the dust disc \citep[e.g.][]{Gaensicke2006, Brinkworth2009, Melis2010}. A widely accepted model is that the white dwarfs are accreting debris of disrupted minor planets that survived the post-main-sequence evolution of the white-dwarf progenitor. These minor planets would have been perturbed into the white dwarf's tidal radius and subsequently disrupted \citep{DebesSigurdsson2002, Jura2003}. Therefore, high-resolution spectroscopic observations of the heavily polluted white dwarfs can uniquely reveal the bulk chemical compositions of these extrasolar minor planets \citep{Zuckerman2007,JuraYoung2014}.

No evidence, however, for such a disintegrating minor planet has ever been directly identified until relatively recently. WD~1145+017 happened to be observed by the K2 mission (in Campaign 1), and it was observed to display transits with multiple periods ranging from 4.5-4.9~hours \citep{Vanderburg2015}. Followup observations came quickly \citep[e.g.][]{Gaensicke2016, Rappaport2016, Gary2017, Croll2017} and found that the transit durations range from $\sim$~3~min to as long as an hour---much longer than expected for a solid body \citep{Vanderburg2015}.  The transits are inferred to be caused by the passage of dust clouds rather than solid bodies.  Presumably each periodicity represents the orbit of a different underlying body that supplies the dusty effluents.  In turn, these bodies are currently hypothesised to be fragments from the tidal disruption of an asteroidal parent body.  At these orbital periods all the orbiting objects lie close to the white dwarf's tidal radius. The transit profiles are variable, asymmetric, and display depths up to 60\% \citep[e.g.][]{Gaensicke2016, Rappaport2016, Gary2017, Croll2017}; they are morphologically similar to transits of disintegrating planets around main-sequence stars \citep[e.g.][]{Rappaport2012}. On a timescale of a few weeks, there can be significant evolution of the transit shape and depth \citep[e.g.][]{Gaensicke2016, Gary2017, Croll2017}. The origin, creation mechanism, and lifetimes of these orbiting bodies are currently uncertain at best, with inferred masses in the range of $10^{17}-10^{24}$~g \citep{Vanderburg2015,Rappaport2016}. Additional dynamical simulations support the statement that the orbiting objects should be no more massive than Ceres \citep{Gurri2017}. In order to produce the observed transit features, the disintegrating objects are likely to be in circular orbits and also differentiated \citep{Veras2017}. 

When observed with the Keck Telescope, WD~1145+017 was found to display photospheric absorption lines from 11 elements heavier then helium, making it one of the most heavily polluted white dwarfs known \citep{Xu2016}. In addition, a uniquely rich set of circumstellar absorption lines from 7 elements was detected. These lines tend to have large velocity dispersions ($\sim$ 300 km s$^{-1}$), and may well be associated with gas orbiting near the white dwarf (i.e., at distances of $\sim$5-15 white dwarf radii; \citealt{Redfield2017, VanderburgRappaport2017}). The circumstellar line profiles have changed significantly since their original discovery \citep{Redfield2017}. In addition, WD~1145+017 displays infrared excess from orbiting dust particles \citep{Vanderburg2015}. 

There have been several attempts with multi-wavelength observations to constrain the properties of the transiting material. If the transits are caused by optically thin dust passing in front of the star, the transit depths should be wavelength dependent due to the wavelength dependence behaviours of Mie scattering cross sections. From simultaneous V- and R-bands observations, \citet{Croll2017} concluded that the particle radii must be larger than 0.15~$\mu$m or smaller than 0.06~$\mu$m. \citet{Alonso2016} reported observations from 4800 to 9200~{\AA} and found no difference in the transit depth across that wavelength range. They concluded that particle sizes less than 0.5~$\mu$m can be excluded for the common minerals. \citet{Zhou2016} extended the observations to J band and still found no wavelength dependence of the transits. They placed a 2~sigma lower limit on the particle size of 0.8~$\mu$m. The first detection of a wavelength-dependent transit was recently reported in \citet{Hallakoun2017}, which features ``bluing" -- transits are shallower in the u' band than those in the r' band, in contrast to the usual expectation in a dusty environment. After exploring different scenarios, they concluded that the most likely explanation is the reduced absorption of circumstellar lines during transits.

In this work we extend the photometric observations to 4.5~$\mu$m. This paper is organised as follows. In Sect.~\ref{sec:observations} we present the details of our observations and the data analysis methods. In Sect.~\ref{sec:colour} we derive transit depth ratios and find that they are consistent at all the observed wavelengths. We interpret this as a result of the prevalence of large grains in Sect.~\ref{sec:Dearth}. We explore a model that would explain the dearth of small grains in Sect.~\ref{sec:model}. The connection between the transiting material and the infrared excess is discussed in Sect.~\ref{sec:IRExcess} and the conclusion is given in Sect.~\ref{sec:conclusion}.

\section{Observations and Data Reduction }
\label{sec:observations}

\begin{figure*}
\includegraphics[width=0.7\textwidth]{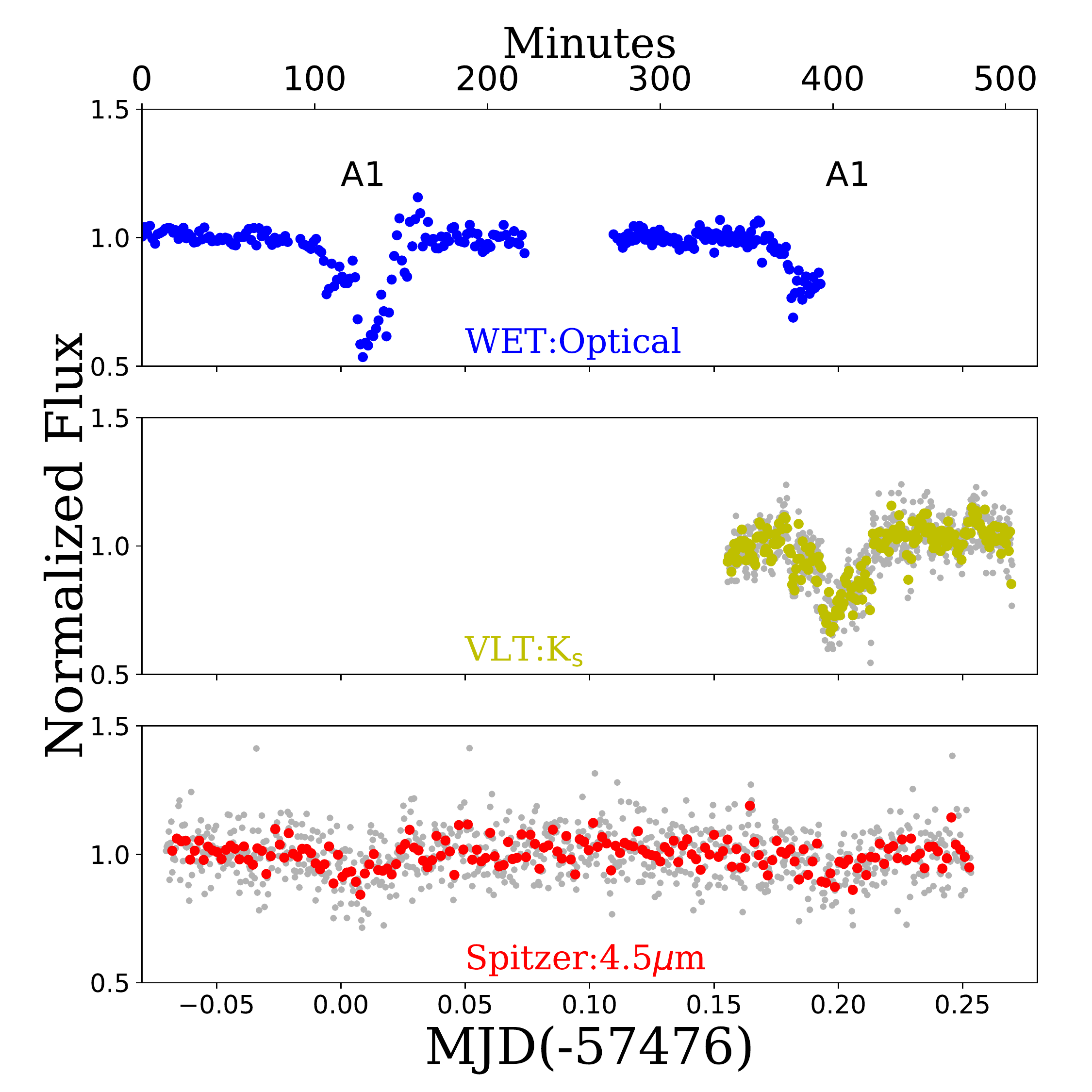}
\caption{The light curves from all the observations on March 28-29, 2016. There was one main transit feature during a full period of 270~min and we denote it as A1. For the K$_\mathrm{s}$ and 4.5~$\mu$m band observations, the grey dots represent individual measurements. The yellow dots are smoothed with every 3 data points and the red dots are smoothed every 5 data points.
\label{Fig:2016LC}}
\end{figure*}

\begin{figure*}
\includegraphics[width=0.7\textwidth]{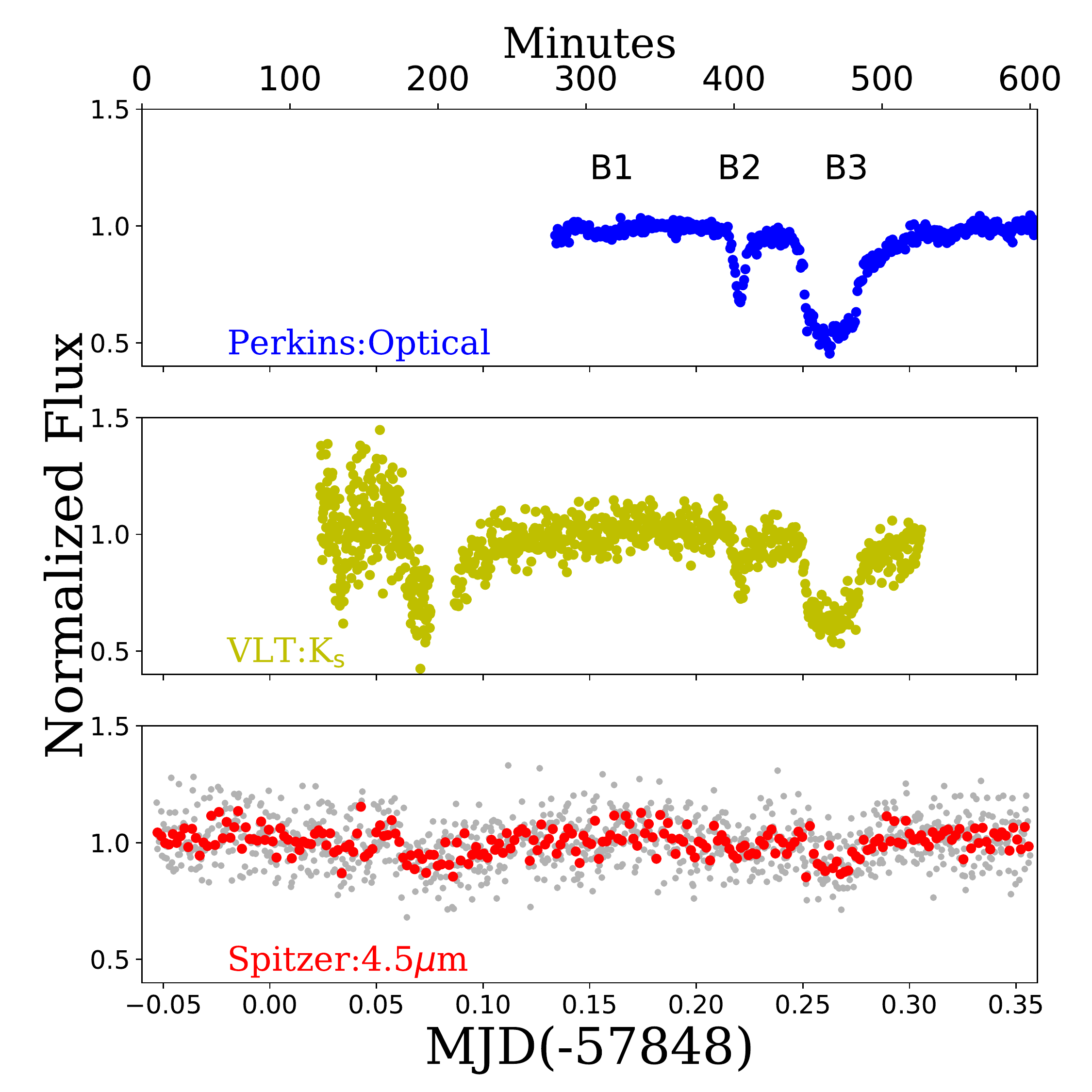}
\caption{The light curves from all the observations on April 4-5, 2017. There are three main transit features in a full period of 270 min, denoted as B1, B2, and B3.  All the notations are the same as those in Fig.~\ref{Fig:2016LC}.
\label{Fig:2017LC}}
\end{figure*}

We have carried out two epochs of multi-wavelength photometric observations that cover optical to 4.5~$\mu$m during both 2016 March 28-29 and 2017 April 4-5. The observing logs are listed in Table \ref{Tab:Observations} and the light curves are shown in Figs. \ref{Fig:2016LC} and \ref{Fig:2017LC}. We describe the observations and data reduction methods in the following section.

\begin{table*}
\caption{Observing logs}
\begin{tabular}{llll}\label{Tab:Observations}
\\
\hline \hline
Instrument & Central Wavelength & Observing Time (UT) & Exposure Time (s) \\
  \hline
  \multicolumn{4}{c}{2016} \\
Meyer	& 	0.48 $\mu$m& Mar 28, 21:00 - Mar 29, 04:37$^a$& 60 	\\
VLT/HAWK-I & 2.1 $\mu$m (K$_\mathrm{s}$) & Mar 29, 03:43 - 06:28	& 15\\
{\it Spitzer}/IRAC	& 4.5 $\mu$m	& Mar 28, 22:18 - Mar 29, 06:04	& 30 \\ 
\\
  \multicolumn{4}{c}{2017}  \\
Perkins/R	& 0.65 $\mu$m	& Apr 5, 03:12 - 09:50 	& 45\\
VLT/HAWKI & 2.1 $\mu$m (K$_\mathrm{s}$) & Apr 5, 00:33 - 07:19	& 15, 30$^b$\\
{\it Spitzer}/IRAC & 4.5 $\mu$m & Apr 4, 22:42 - Apr 5, 08:33	& 30 \\
\hline
\end{tabular}
\\
{\bf Note.} \\
$^a$There are some gaps in the light curve due to passing clouds.\\
$^b$The first set of observations ($\sim$ 75 min) was executed in 15 s exposure time. \\
\end{table*}

\subsection{WET \& Perkins: Optical}
\label{sec:OpticalObs}

For the optical observation in 2016, we used the 0.6m Paul \& Jane Meyer Observatory Telescope in the Whole Earth Telescope (WET) network with the BG~40 filter. Data reduction was performed following the procedures developed for the WET observations \citep{Provencal2012}. There were passing (terrestrial) clouds, which caused some gaps and increased scatter in certain portion of the light curve, as shown in Fig.~\ref{Fig:2016LC}.

During the 2017 observing run, we arranged observations with several optical telescopes. The details of these observations are described in Appendix \ref{sec:app}. Here, we focus on data taken on the 1.8m Perkins telescope at the Lowell Observatory using the Perkins Re-Imaging System (PRISM, \citealt{Janes2004}), which has the best data quality. The observation was designed following previous WD~1145+017 observations \citep{Croll2017}. The observing conditions were moderately good. Seeing was around 3{\farcs}0 and there were thin cirrus clouds at the beginning of the observation, which developed substantially throughout the night. Data reduction was performed following a custom pipeline \citep{DalbaMuirhead2016,Dalba2017}. 

\subsection{VLT/HAWKI: K$_\mathrm{s}$ \label{sec: LC_VLT}}

On 2016 March 28-29, WD~1145+017 was monitored in $K_\mathrm{s}$ band with the HAWK-I (High Acuity Wide field K-band Imager; \citealt{Pirard2004}) at the Very Large Telescope (VLT). The camera is equipped with four HAWAII 2RG 2048$\times$2048 detectors, with a plate scale of $\sim$0{\farcs}106 pixel$^{-1}$. 
We used the {\it Fast Jitter} mode that allows one to window down the detectors, greatly reducing the readout overheads. Following similar procedures outlined in \citet{Caceres2011}, we applied windows of 128$\times$256\,pixels per detector stripe. We recorded a sequence of blocks, each consisting of 60 exposures with 15 s integration time. Throughout the observations, we kept the nearby star ULAS\,J114829.42+012707.6 
($K_\mathrm{s}$=16.678$\pm$0.042)
on the same detector as the science target to provide a flux reference. The weather conditions were moderate and there were thin clouds passing by during the observations. The DIMM seeing was about 0{\farcs}55 in the optical. 

\begin{figure}
\includegraphics[width=\columnwidth]{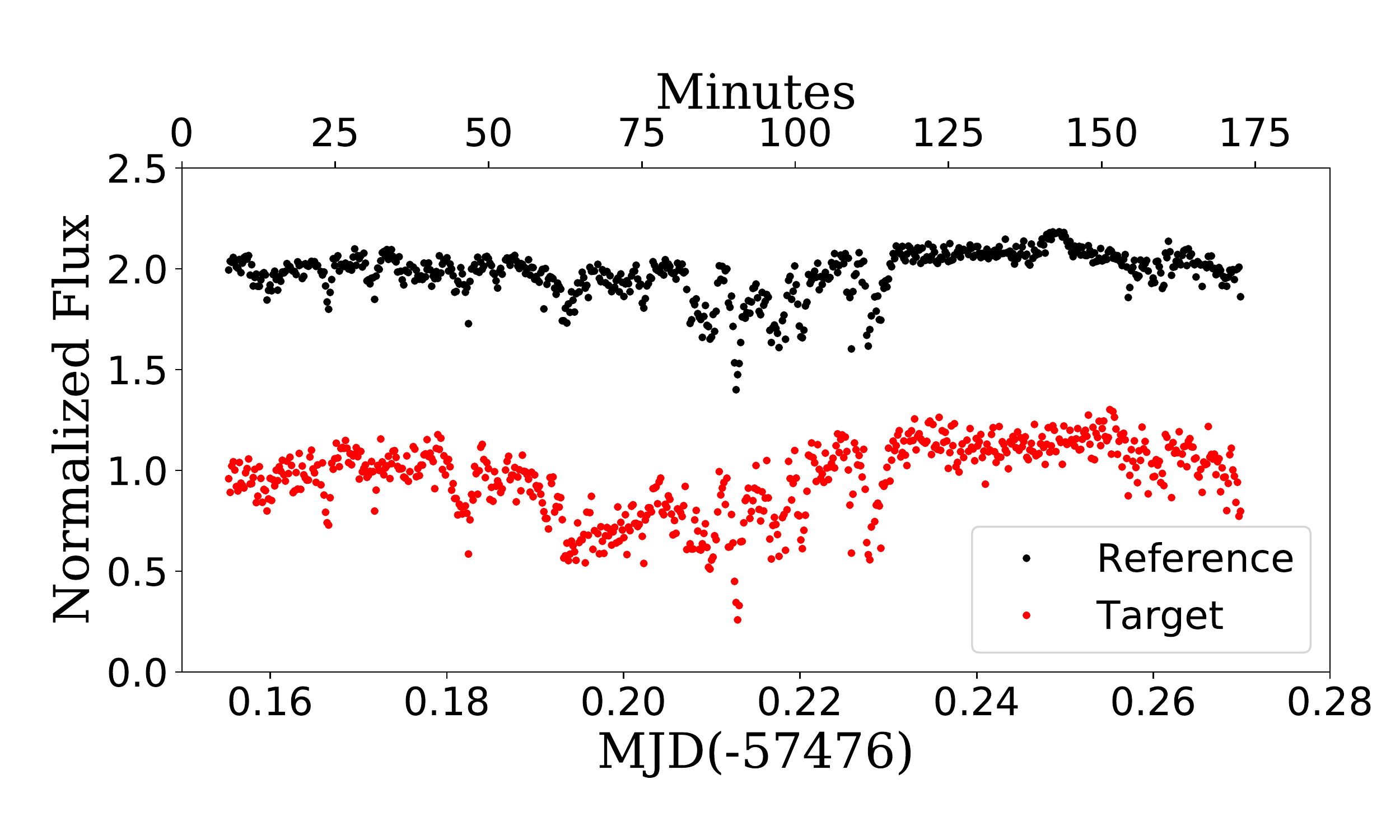}
\includegraphics[width=\columnwidth]{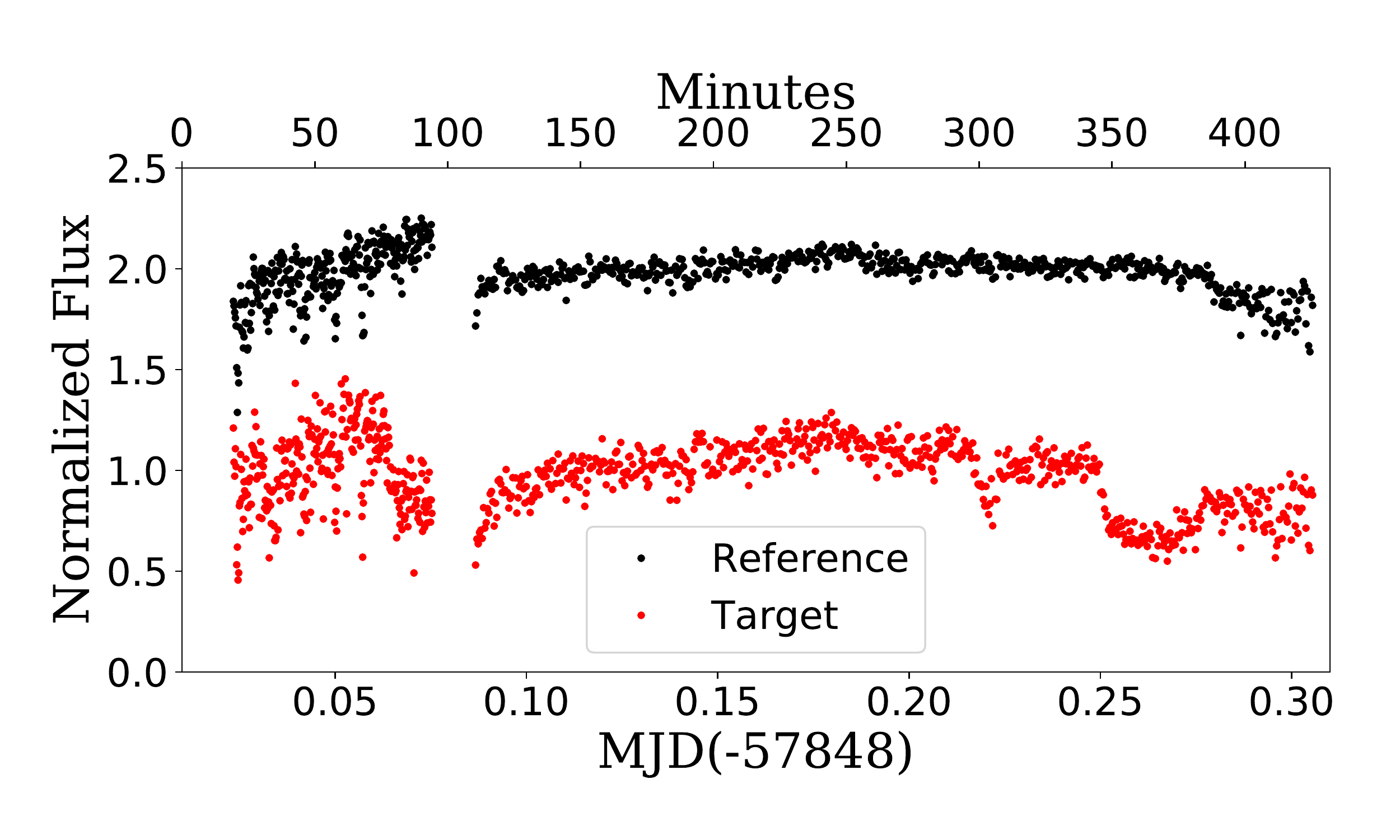}
\caption{K$_\mathrm{s}$ band light curve of WD~1145+017 from the HAWK-I on the VLT. Red dots are the normalised flux for our target WD~1145+017 while black dots are for the reference star ULAS\,J114829.42+012707.6. For clarity, the black dots are offset by 1.0 from the red dots. The top panel is for the 2016 dataset, during mediocre weather conditions with passing clouds; the bottom panel is for the 2017 dataset with good weather conditions.
\label{Fig: HAWKI}}
\end{figure}

Individual images were dark-subtracted and flat-field corrected with sky flats taken immediately after the observation. Aperture photometry was performed with aperture radii of  5, 7, and 10~pixels. The sky background was estimated from an annulus of between 10 and 20~pixels. We found that the light curve from the 7~pixel (0{\farcs}75) radius aperture has the best quality and use this light curve for the rest of our analysis. Both the target light curve and the reference star light curve were normalised by dividing by a constant, as shown in Fig. \ref{Fig: HAWKI}.

We repeated the observations with HAWK-I on April~4-5, 2017 with a similar set-up. The observing conditions were better: clear sky with optical seeing at the beginning of 1{\farcs}0, which decreased to 0{\farcs}5 toward the end of the observation. We started the observations with a block of 60 exposures with 15 s integration each. We repeated this block five times before noticing that the target was not visible in the guider. Out of concern that the target had drifted out of the small field of view, we stopped the sequence, reacquired the target, and increased the exposure time to 30\,s. After that, we observed in blocks consisting of 30 exposures with 30-s integration times until the end of the observation.

Data reduction was performed following the same method as for the 2016 dataset. We adopted an aperture radius of 8 pixels (0{\farcs}85), and a sky annulus between 10 and 20 pixels for aperture photometry. In retrospect, our target did not disappear around 57848.08 (MJD); in fact, there was a deep transit, which made the target difficult to see in individual images. Because the data quality was better with 30-s exposure time, we focus on these for the following analysis.

\subsection{Spitzer/IRAC: 4.5~$\mu$m \label{sec: LC_Spitzer}}

We were awarded time with the Infrared Array Camera (IRAC; \citealt{Fazio2004}) on the {\it Spitzer Space Telescope} under a DDT program to observe WD~1145+017 simultaneously with the VLT on March 28-29, 2016. The observation was designed following ``Advice for designing high precision photometry observations"\footnote{\href{http://irachpp.spitzer.caltech.edu/page/Obs\%20Planning}{http://irachpp.spitzer.caltech.edu/page/Obs\%20Planning}}. The science observations consisted of 900 exposures of 30 s each in the 4.5 $\mu$m filter in stare mode. The readout mode was full array. The target was put in the well-characterised pixel (``sweet spot"). Before the science observation, we arranged a 30-minute exposure to eliminate the initial drift of the instrument \citep{Grillmair2012}. We also included a 10-minute post-observation with the same setup as the pre-observation. 

For the data reduction, we started with the CBCD (Corrected Basic Calibrated Data) files, which are flux-calibrated and artefact-corrected files from the pipeline (IRAC Instrument Handbook). We excluded a few frames when there was a cosmic ray close to the target. The pre-observation and post-observation frames were median combined and smoothed to create a localised dark frame, which was then subtracted from all CBCD files to remove any residual patterns. To produce the light curve, we used codes that are publicly available for IRAC high precision photometry\footnote{\href{http://irachpp.spitzer.caltech.edu/page/contrib}{http://irachpp.spitzer.caltech.edu/page/contrib}}. The \textsc{IDL} program \textsc{box$\_$centroider.pro} was used to locate the position of our target in each exposure. We found the average target position to be X = 24.08 $\pm$ 0.07, Y = 232.18 $\pm$ 0.08, which is very close to the sweet spot at X=24.0, Y=232.0. These positions were used as input parameters for aperture photometry in \textsc{aper.pro}, where an aperture radius of 2~pixels and a sky annulus of 12-20~pixels were used. The last step was to correct for the intrapixel gain with a pre-gridded pixel map in \textsc{iracpc$\_$pmap$\_$corr.pro}. The normalization was done by dividing the light curve by a constant.

The same set of observations was repeated on April 4-5, 2017. A total of 1140 exposures for science observations were obtained with 30\,s exposure time each. The observation covered more than two full periods. The average centroid position for the target was 24.08 $\pm$ 0.08 in X and 232.31 $\pm$ 0.09 in Y. We followed the same data reduction procedures as for the 2016 dataset.

In order to measure the absolute flux in 4.5 $\mu$m, we took the median value of the measured out-of-transit flux. After aperture correction, we found that the flux was 55.0 $\pm$ 2.3 $\mu$Jy and 55.5 $\pm$ 2.5 $\mu$Jy in 2016 and 2017, respectively. Evidently, the infrared excess remained stable. We took the average of these numbers and also included a conservative 5\% calibration uncertainty for IRAC, as found by previous dusty white dwarf studies \citep{Jura2007b}. The final flux is 55.0 $\pm$ 3.2 $\mu$Jy at 4.5 $\mu$m, which is consistent with the WISE W-2 flux of 43~$\pm$~14~$\mu$Jy. 

\section{Transit Depths in Different Wavelengths}
\label{sec:colour}

\subsection{The Observed Transit Depth \label{sec: FitDepth}}

In this section, we compare the transit depths at the different observed wavelengths. Some of the observations cover more than one full orbital period, so we folded the light curve about a period of 269.47~min \citep{Gary2017}. This phase folding is unlikely to affect our analysis because typically, there is little evolution in the light curve on an orbital time scale (e.g. \citealt{Gary2017}). We confirmed this with our April 2017 ground-based observations, which spanned the full $\sim$~10h {\it Spitzer} observations and showed only negligible evolution of the transit shapes and depths over the course of the observations (see Appendix \ref{sec:app}).

For both the 2016 and 2017 datasets, we start with the optical light curve, which has the lowest point-to-point scatter. Following previous work of fitting asymmetric transit features \citep[e.g.][]{Rappaport2014,Zhou2016,Croll2017,Gary2017}, we model the transits of WD~1145+017 as a sum of  asymmetric hyperbolic secant (AHS) functions:
\begin{equation}
f(p)=f_{\mathrm{0}} \, \left[1 - f_{\mathrm{dip}}(p)\right] ~{\equiv}~ f_\mathrm{0} \, \left(1 - \sum \limits_{i}  \frac{2f_\mathrm{i}}{e^\frac{p-p_i}{\phi_{i1}}+e^{-\frac{p-pi}{\phi_{i2}}}} \right)  
\label{Equ:AHS}
\end{equation}  
where $f(p)$ represents the normalised light curve, $f_\mathrm{0}$ fits the continuum level, and $f_\mathrm{dip}(p)$ is the fractional flux change during the asymmetric transits. In the second term, $i$ represents the number of AHS components needed to provide a good fit to the transit profile (typically $i = 2-4$), $p_i$ represents a phase near the deepest point during a transit, $\phi_{i1}$ and $\phi_{i2}$ represent the ingress and egress duration of a transit, respectively. 

Assuming the light curves at all the wavelengths have the same shape but different transit depths, we can fit all the light curves with the following form,
\begin{equation}
f_\mathrm{\lambda}(p)=f_\mathrm{\lambda,0} \, \left[1 - D_\mathrm{\lambda} \, f_\mathrm{dip}(p) \right]   
\label{Equ:D}
\end{equation}
There are only two free parameters here, i.e. $f_\mathrm{\lambda,0}$ and $D_\mathrm{\lambda} $, while the parameters for $f_\mathrm{dip}(p)$ can be taken from the best-fit parameters for the optical light curve. $f_\mathrm{\lambda,0}$ is used to fit the continuum level and $D_\mathrm{\lambda} $ is the observed transit depth ratio between the wavelength of interest relative to optical. 

Our fitting method adopts a Levenberg-Marquardt algorithm to derive a least square fit, and the uncertainty comes from the covariance matrix of the best fit values. To estimate the uncertainty in the depth of transits in the optical light curve $D_\mathrm{opt}$, we repeated the fit following Eq. \eqref{Equ:D}. The observed uncertainty for the transit depth ratios between a given wavelength $\lambda$ and the optical includes the uncertainty from $D_\mathrm{opt}$ and the uncertainty from $D_\lambda$. 

\subsubsection{2016 Dataset}

\begin{figure*}
\includegraphics[width=0.7\textwidth]{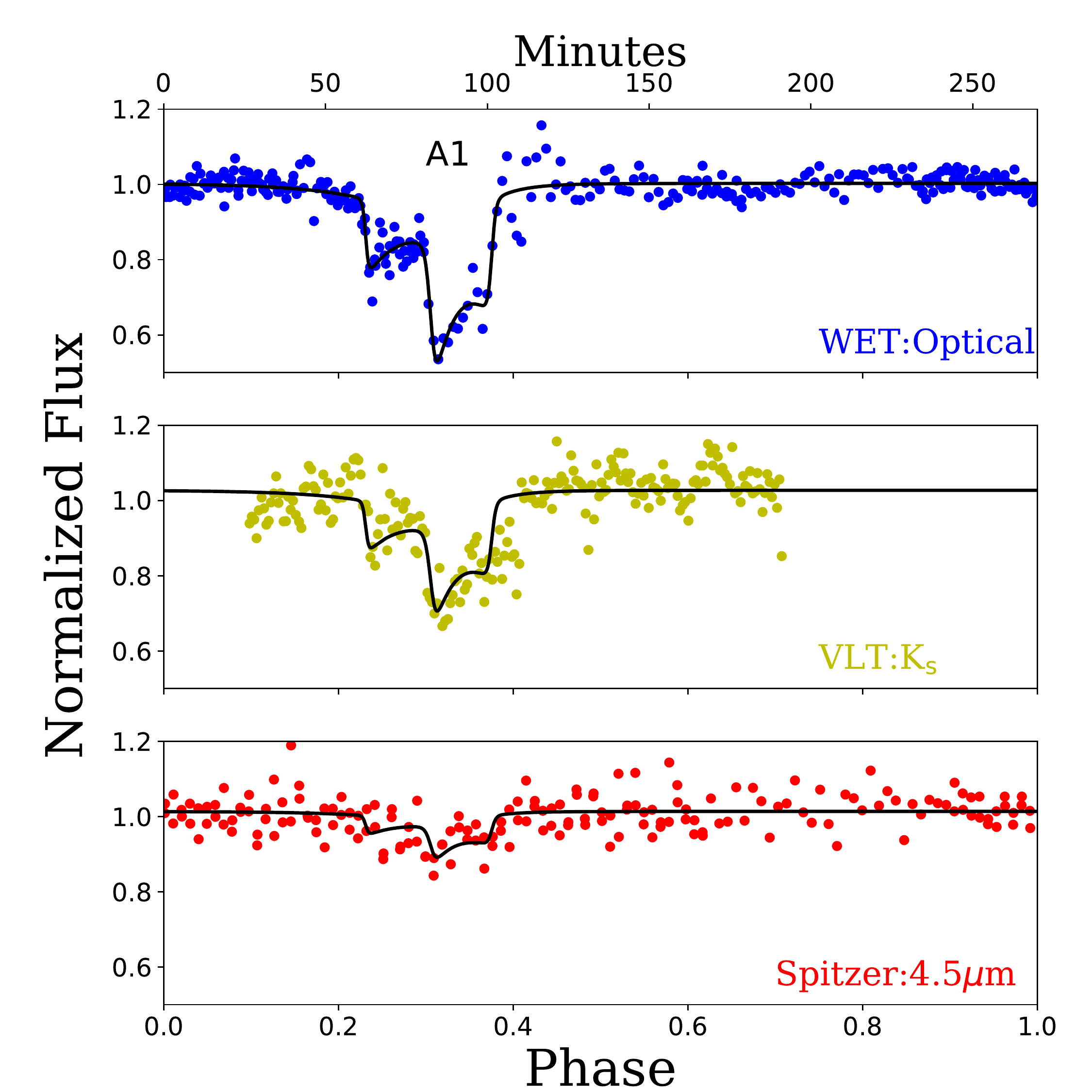}
\caption{{Phase-folded transit light curve in the optical, K$_\mathrm{s}$ band, and 4.5 $\mu$m, respectively on March 28-29, 2016.} The black line represents the best-fit models with parameters listed in Tables \ref{Tab:AHSFit} and \ref{Tab:Depth}.
\label{Fig: LC_Fit_2016}}
\end{figure*}

We set the reference time as 57475.95 (MJD) and the phase-folded light curves are shown in Fig. \ref{Fig: LC_Fit_2016}. During our observation, there was one main transit feature with a complex shape. We refer to this feature as A1. It was first detected on January 21, 2016 (denoted as `G6121' in \citealt{Gary2017} and `A1' in \citealt{Hallakoun2017}). This feature was well covered in the optical, K$_\mathrm{s}$ band, 4.5~$\mu$m. 

For the WET light curve, we used three AHS components for A1 following Eq.~\eqref{Equ:AHS} and the best-fit parameters are listed in Table \ref{Tab:AHSFit}. We fitted all the light curves with the same functional form but different transit depth $D$ with Eq.~\eqref{Equ:D}, with the results listed in Table \ref{Tab:Depth}. The observed transit depths are different at all three wavelengths.

\begin{table*}
\begin{center}
\caption{Best-fit parameters for the AHS components of different dip features in the optical band}
\begin{tabular}{lllllllll}\label{Tab:AHSFit}
\\
\hline \hline
Dip &f$_\mathrm{0}$ &  i	&f$_\mathrm{i}$	& p$_i$	& $\phi_{i1}$	& $\phi_{i2}$	\\
\hline
A1	&1.003 $\pm$ 0.003	& 3		& 0.105 $\pm$ 0.012& 0.231 $\pm$ 0.001	&0.048 $\pm$ 0.011 &0.002 $\pm$ 0.001	 \\
	&				&		& 0.214 $\pm$ 0.025& 0.306 $\pm$ 0.002&0.028 $\pm$ 0.007 & 0.003 $\pm$ 0.001	\\
	&				&		& 0.151 $\pm$ 0.023& 0.375 $\pm$ 0.001&0.002 $\pm$ 0.001 & 0.073 $\pm$ 0.010\\
\\
B1	&  1.001 $\pm$ 0.002	&1  & 0.027 $\pm$ 0.006	& 0.147 $\pm$ 0.003 & 0.003 $\pm$ 0.002 & 0.042 $\pm$ 0.013\\
\\
B2	& 0.979 $\pm$ 0.008	& 1 	& 0.284 $\pm$ 0.026	&  0.456 $\pm$ 0.003 & 0.015 $\pm$ 0.004 & 0.008 $\pm$ 0.002\\
\\
B3	& 0.979 $\pm$ 0.008	& 2	& 0.166 $\pm$ 0.022	& 0.622 $\pm$ 0.002 & 0.130 $\pm$ 0.017 & 0.004 $\pm$ 0.002\\
	&	&	& 	 0.276 $\pm$ 0.035	& 0.729 $\pm$ 0.009 & 0.027 $\pm$ 0.006 & 0.064 $\pm$ 0.010 \\
\hline
\end{tabular}
\end{center}
{\bf Note.} \\
The parameters are defined in Eq. \eqref{Equ:AHS}. We considered phase 0-0.3 for B1, 0.3-0.55 for B2, and 0.55-1.0 for B3. For different dips, $f_0$ is slightly different because of imperfect continuum normalisation. 
\\
\end{table*}

\begin{table*}
\begin{center}
\caption{Transit depth ratios at the observed wavelengths}
\begin{tabular}{lcccccccc}\label{Tab:Depth} \\
\hline \hline
& D$_\mathrm{opt}$	 & D$_\mathrm{K_\mathrm{s}}$ & D$_\mathrm{K_\mathrm{s}}$/D$_\mathrm{opt}$ & D$_\mathrm{4.5 \mu m}$ & D$_\mathrm{4.5 \mu m}$/D$_\mathrm{opt}$\\
\hline
A1 	& 1.000 $\pm$ 0.021	 	& 0.664 $\pm$ 0.034 & 0.664 $\pm$ 0.037 & 0.256 $\pm$ 0.032 & 0.256 $\pm$ 0.032 \\
B2	 & 1.000 $\pm$ 0.056	& 0.768 $\pm$ 0.057	& 0.768 $\pm$ 0.071 & 0.309 $\pm$ 0.073 &  0.309 $\pm$ 0.075\\
B3	 & 1.005 $\pm$ 0.013	& 0.796 $\pm$ 0.018	& 0.792 $\pm$ 0.021 & 0.240 $\pm$ 0.026 & 0.239 $\pm$ 0.026\\
\\
Average & 	& &	0.741 $\pm$ 0.028	&	& 0.268 $\pm$ 0.029\\
\hline
\end{tabular}
\end{center}
{\bf Note.}\\
D is the best-fit parameter defined in Eq. \eqref{Equ:D}. $D_\mathrm{opt}$ is not exactly unity because it depends on the range chosen to calculate the out-of-transit flux, which is different for the B2 and B3 dips. The uncertainty in $D_\mathrm{opt}$ illustrates a minimum uncertainty even when fitting the same data in different ways (with Eq.~\eqref{Equ:AHS} and Eq.~\eqref{Equ:D}).
\\
\end{table*}

\subsubsection{2017 Dataset}

\begin{figure*}
\includegraphics[width=0.7\textwidth]{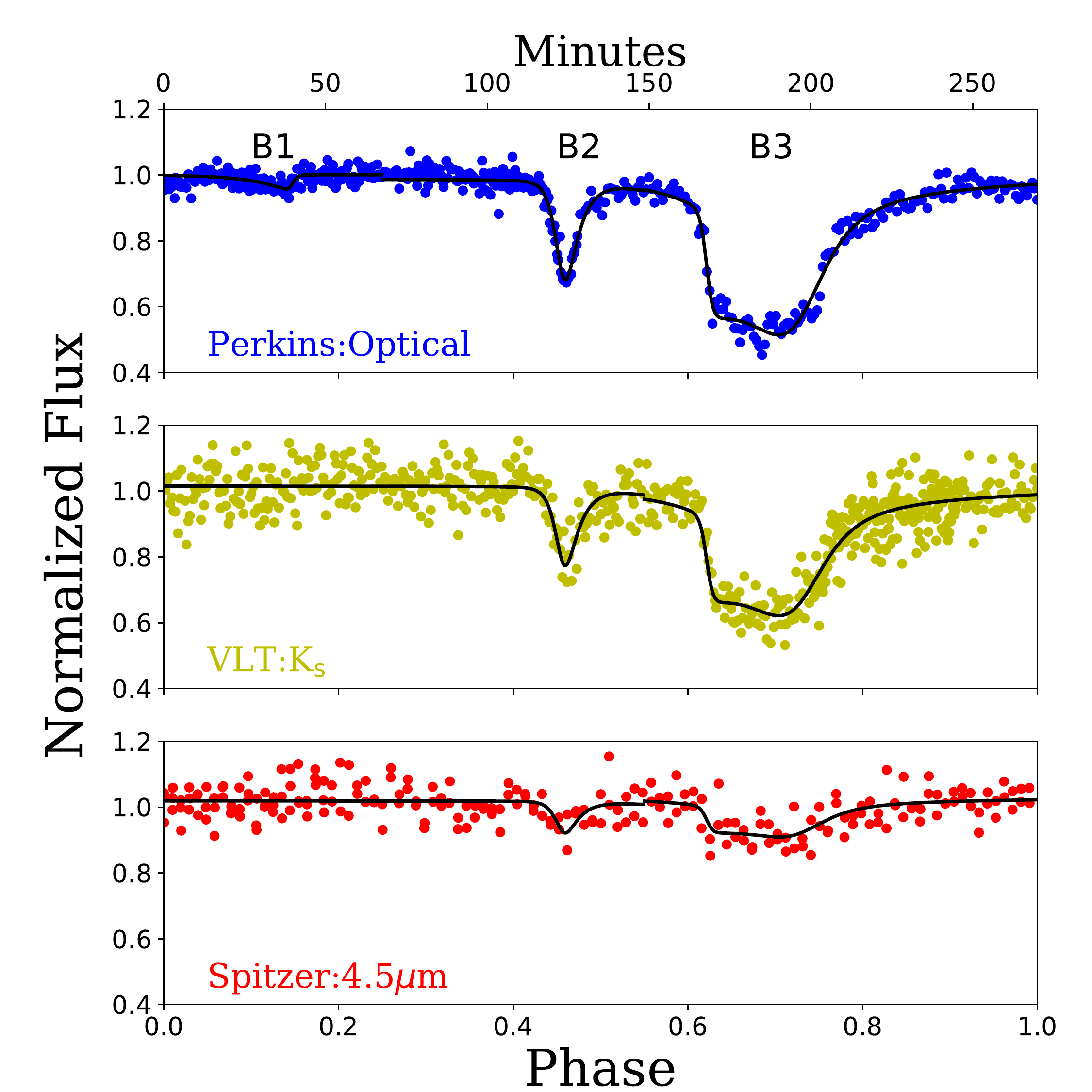}
\caption{Similar to Fig. \ref{Fig: LC_Fit_2016} except for the 2017 dataset.
\label{Fig: LC_Fit_2017}}
\end{figure*}

We set the reference time as 57848.13 (MJD) and the phase-folded light curves are shown in Fig. \ref{Fig: LC_Fit_2017}. There were three main transit features during a full orbital period, denoted as B1, B2, and B3 (see Appendix~\ref{sec:app} for details). The B1 Dip was shallow and not detected in the K$_\mathrm{s}$ band nor the {\it Spitzer} band, due to their relatively low signal-to-noise ratios compared to the optical light curve. Here, we focus on transits B2 and B3, which are better suited for studying the wavelength-dependence of the transits. We believe B2 was due to the same orbiting body that produced A1 in 2016 and B3 could be related to some other features observed in the previous season as well \citep{Rappaport2017}. 

We followed the same analysis procedures outlined in Sect.~\ref{sec: FitDepth}. We started with the Perkins light curve and used the functional form for $F_{\rm dip}(p)$ to fit the light curves at all the wavelengths. The results are listed in Tables~\ref{Tab:AHSFit} and \ref{Tab:Depth}. The values for B2 and B3 are comparable, while the overall uncertainties for B3 are smaller because the transit was deeper and lasted longer. 

\subsection{Transit Depth Correction for Dust Emission} 

In both datasets, we find that the transit depths of the observed flux are different at all the wavelengths. It is deepest in the optical and shallower at longer wavelengths. There is no change in the transit depth ratios between these two epochs. 

Because WD~1145+017 has an infrared excess starting from the K$_\mathrm{s}$ band \citep{Vanderburg2015}, this will necessarily dilute the transit signal. To derive the transit depth ratio corrected for the dust emission, we need to use the intrinsic white dwarf flux rather than the measured flux. As a result, the corrected transit depth ratio can be calculated as:
\begin{equation}
D_\mathrm{corr}= \frac{F_\mathrm{obs}}{F_\star} \times D
\label{Equ:1/f}
\end{equation}
where F$_\mathrm{obs}$ and F$_\star$ represents the measured out-of-transit flux and the expected flux from the white dwarf, respectively. We know that F$_\mathrm{obs}$ is constant during the two epochs of our observations. Based on the colour of WD~1145+017, we find that F$_{\star}$ has little extinction from either circumstellar or interstellar material and therefore F$_{\star}$ can be derived from white dwarf model calculations. We take the correction factor $F_\mathrm{obs} / F_\star$ to be a constant.

We calculated white dwarf model spectra with parameters listed in Table \ref{Tab:WDParameters} and derived the fact that the white dwarf flux is 53.5~$\mu$Jy at K$_\mathrm{s}$ band and 13.0~$\mu$Jy at 4.5~$\mu$m. Varying the temperature by 500~K and log g by 0.2~dex, we found the white dwarf flux could change by at most 2\% and we adopted that as the uncertainty of the intrinsic white dwarf flux. The measured total flux is 69.4~$\pm$~5.2~$\mu$Jy in K$_\mathrm{s}$ band from the UKIDSS and 55.0~$\pm$~3.2~$\mu$Jy at 4.5~$\mu$m, as derived in Sect. \ref{sec: LC_Spitzer}. Therefore, the correction factors are
$$\left( \frac{F_\mathrm{obs}}{F_\star} \right)_{\mathrm{K_{s}}} = 1.30\pm 0.10 $$
$$\left( \frac{F_\mathrm{obs}}{F_\star} \right)_{4.5 \mu m} = 4.23 \pm 0.26.$$ 
Following Equ.~\eqref{Equ:1/f}, the corrected average transit depth ratios are:
$$ \left( \frac{D_\mathrm{K_\mathrm{s}}}{D_\mathrm{opt}} \right)_\mathrm{corr} = 0.96 \pm 0.08 $$
$$ \left( \frac{D_{4.5 \mu m}}{D_\mathrm{opt}} \right)_\mathrm{corr} = 1.13 \pm 0.14 $$ 

\begin{table*}
\centering
  \caption{WD~1145+017 system parameters}
  \label{Tab:WDParameters}
  \begin{tabular}{lccc}
  \hline
  \hline
  Parameter & Symbol & Value & Reference \\
  \hline
  WD \\
  Effective Temperature & $ T_\mathrm{\star} $ & $ 15{,}900 $\,K & \citet{Vanderburg2015} \\
  Surface Gravity & $ \log g $ & $ 8.0 $ &  \citet{Vanderburg2015} \\
  Distance	& d	& 174 pc	&  \citet{Vanderburg2015} \\
  Mass & $ M_\mathrm{\star} $ & $ 0.6 $\,M$_\mathrm{\odot} $ & \citet{Dufour2017} \\
  Radius & $ R_\mathrm{\star} $ & $ 0.013 $\,R$_\mathrm{\odot} $ & \citet{Dufour2017} \\
  \hline
  Transiting Material \\
  Orbital Period & $ P $ & $ 269.47 $\,min & \citet{Gary2017} \\
  Orbital Distance & $ r $ & $ 1.16 $\,R$_\mathrm{\odot} $ & Kepler's third law \\
  \hline
  \end{tabular}
\end{table*}

The corrected transit depths in K$_\mathrm{s}$ band and 4.5 $\mu$m are unexpectedly and even surprisingly consistent with the transit depth in the optical, at least to within the uncertainties. This result tends to indicate that the extinction cross section is independent of wavelength for the observations from optical to 4.5~$\mu$m. 

The main source of uncertainty for this analysis comes from the correction factor $F_\mathrm{obs}/F_\star $ in Eq.~\eqref{Equ:1/f}, which is mostly from the uncertainty of the absolute flux measurements. Currently, it is 7.5\% in K$_\mathrm{s}$ and 5.8\% at 4.5~$\mu$m. Future observations in the infrared will improve the flux measurement in K$_\mathrm{s}$ band. However, IRAC observations will always be limited by its absolute flux calibration, which is $\sim$~5\%.

In the following section, we explore possible physical reasons for the wavelength-independence of the transit depth.

\section{Constraints on the particle size}
\label{sec:Dearth}

The lack of a wavelength dependence of the transit depths could be expected if the dust clouds are optically thick. However, in order to explain both the transit depth and transit duration (e.g. the $\sim$20\% deep and $\sim$90~min long transit reported in \citealt{Alonso2016}), an opaque cloud needs to be both flat and also almost perfectly aligned with the orbital direction. We therefore consider opaque clouds to be an unlikely explanation for the transiting material around WD~1145+017.

An alternative explanation for the colour-independent transit depths is that large grains (i.e., $\gtrsim 1-2 \, \mu$m) dominate the extinction cross-section of the clouds and their cross-sections are nearly independent of wavelength out to 4.5~$\mu$m. 

\begin{figure}
\includegraphics[width=\linewidth]{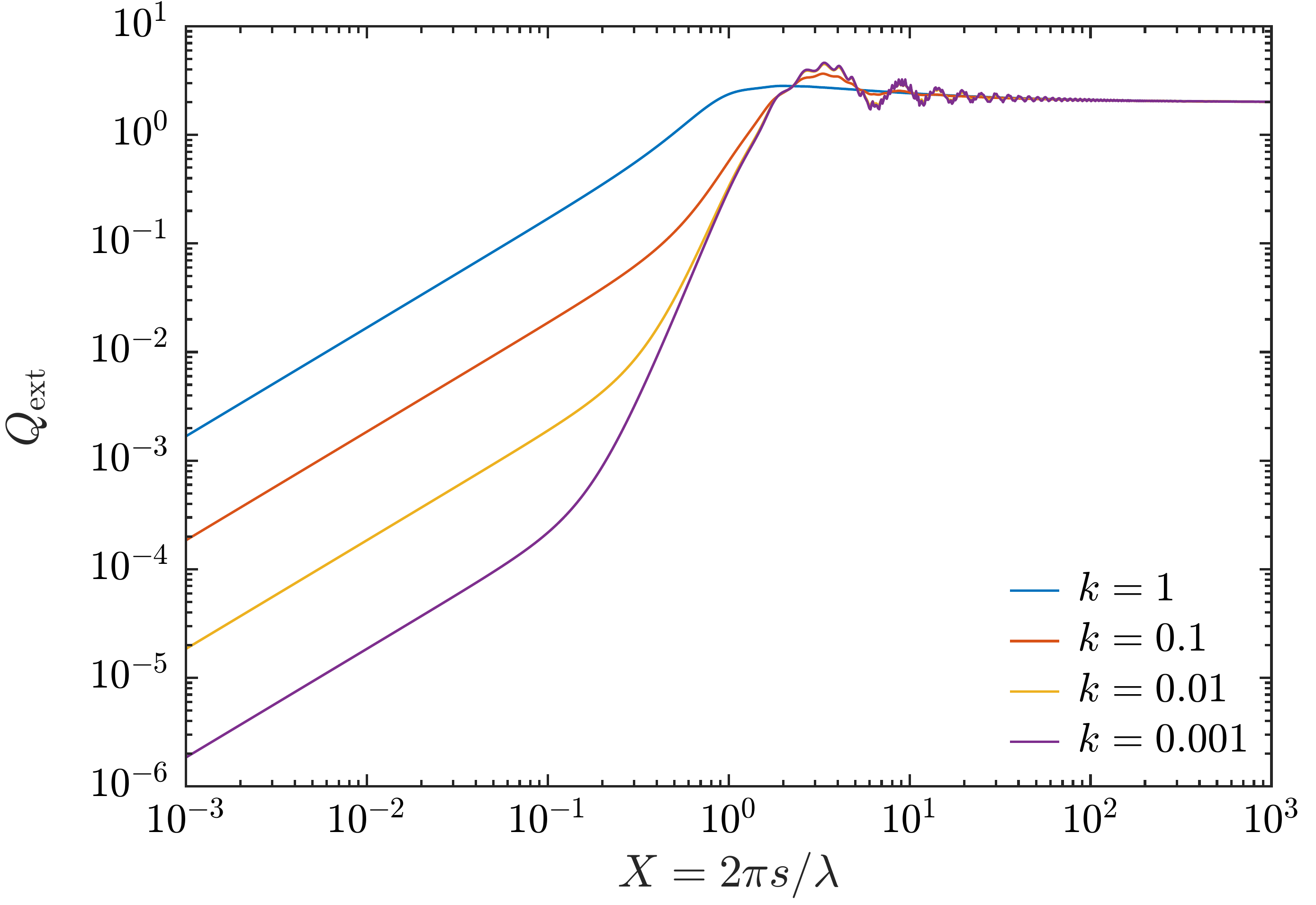}
\caption{Extinction efficiency as a function of scaled particle size $X$ for a range of grains with the real part of the index of refraction n being fixed at 1.6 and the imaginary part k between 0 and 1. For grains that satisfy $X$~$\gtrsim$~2, $Q_{\rm ext}$ is independent of wavelength and remains a constant.}
  \label{Fig:qext}
\end{figure}

We explore the wavelength dependence of the extinction efficiency (the ratio of the extinction cross section $\sigma_{\rm ext}(X)$ to the geometric cross section), $Q_{\rm ext} \equiv \sigma_{\rm ext}(X)/\pi s^2$, where $s$ is the particle radius, $\lambda$ is the observing wavelength, and $X \equiv 2 \pi s/\lambda$, the scaled particle size\footnote{We note that this simple functional form for $Q_{\rm ext}$ can only be used when the imaginary part of the complex index of refraction is essentially independent of wavelength.}. We adopted the Mie algorithm presented in \citet{BohrenHuffman1983} and the results for a range of generic materials are shown in Fig.~\ref{Fig:qext}. Similar to results from previous studies \citep[e.g.][]{Croll2014}, we found that for very small values of $X (\lesssim 0.1)$, $Q_{\rm ext} \propto \lambda^{-1}$ is valid, while for large $X \gtrsim 2$, $Q_{\rm ext} \rightarrow 2$, i.e., independent of $\lambda$. For intermediate values of $X$, $Q_{\rm ext} \propto \lambda^{-4}$.  

Since we find a colourless transit depth between $\lambda \simeq 0.5~\mu$m to 4.5~$\mu$m, we can reasonably infer that $X \gtrsim 2$ even at the longest wavelength of our observations.  Specifically, we find that 
\begin{equation}
X=\frac{2 \pi s}{\lambda} \gtrsim 2
\end{equation}
Therefore, we tentatively conclude that our non-detection of wavelength-dependent transit depths from optical to 4.5 $\mu$m implies that the transiting material around WD 1145+017 must consist of grains whose sizes are largely $\gtrsim$ 1.5~$\mu$m. 

We can also compare the transit depth ratios at different wavelengths with the corresponding Mie extinction cross sections. We consider two generic grain materials (with $n = 1.6$ and k = 0.1, 0.01), and two particle size distributions. 

{\it Hansen Distribution}: For a range of characteristic particle sizes of $\bar{s} = 0.2, 0.5, 1, 2, 5, 10$ $\mu$m, the specific form of the distribution is \citep{HansenTravis1974}: 
\begin{equation} 
n(s) = C s^{(1-3b)/b} e^{-s/ \bar s b} \nonumber
\end{equation} 
where $C$ is a normalization constant, $s$ is the particle radius, and $b$ is the dimensionless variance of the distribution. Following \citet{Zhou2016}, we choose a value for $b$ of 0.1, which provides a distribution that ranges from a factor of roughly $\sqrt{0.1}$ below $\bar s$ to a factor of $\sqrt{10}$ above $\bar s$. The normalised distribution is then 
\begin{equation} 
n(s) = \frac{1}{\Gamma(10) (0.1\bar s)^{10}} \,s^{7} e^{-10 s/ \bar s} \nonumber
\end{equation} 

{\it Cut-on Power-Law Size Distribution}: we also consider a classical power law distribution from a collisional cascade:
\begin{equation}
n(s)=C s^{-3.5}
\label{eq:cuton}
\end{equation}
where the minimum grain size $s_{\rm min}$ is fixed. This distribution is probably a more realistic representation of the particle size distribution as small grains may sublimate rapidly, as discussed in the next section.  Note that in the cut-on power-law size distribution the effective particle size is larger than $s_{\rm min}$.

\begin{figure*}
    \begin{minipage}{0.5\textwidth}
        \includegraphics[width=0.95\textwidth]{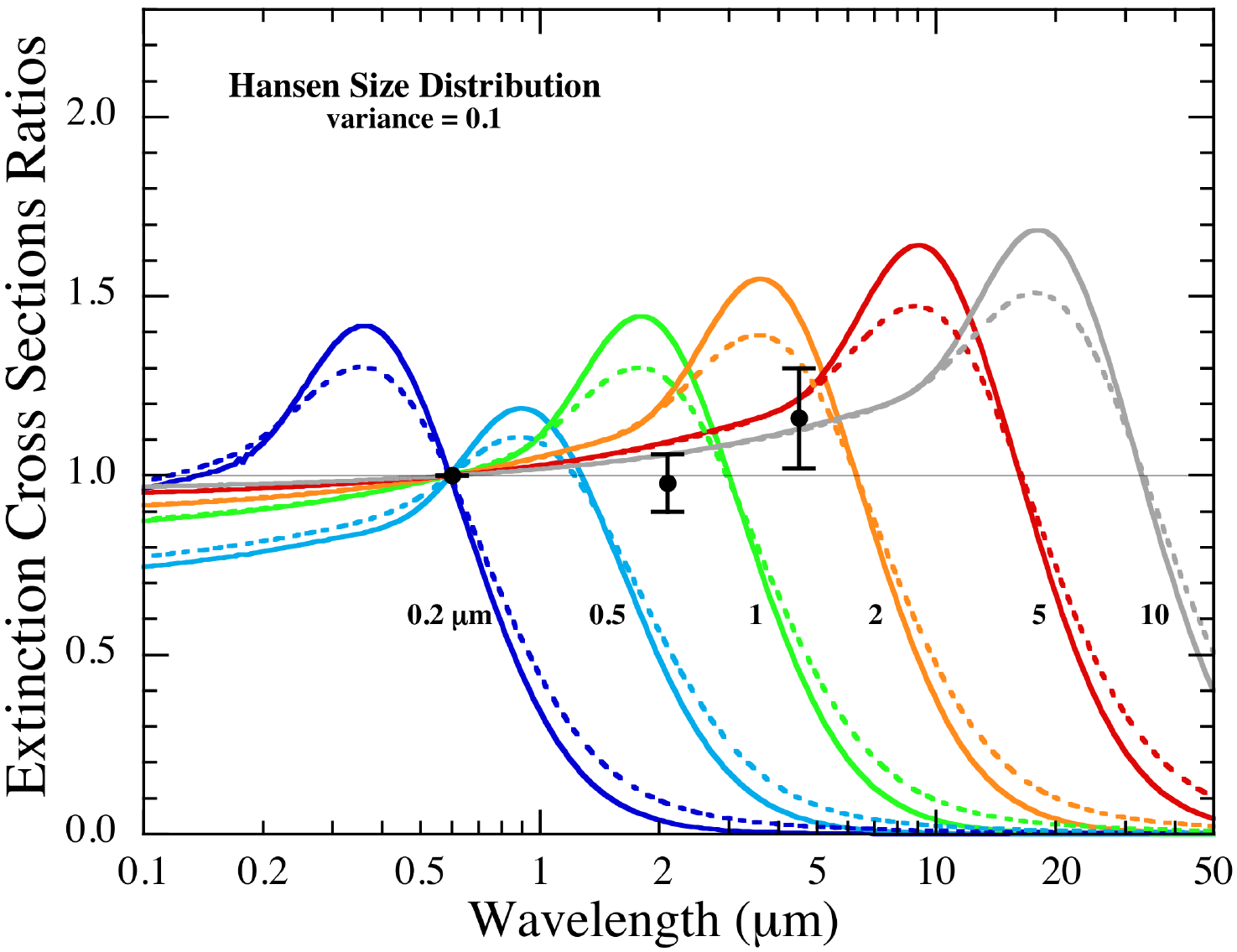} 
    \end{minipage}\hfill
    \begin{minipage}{0.5\textwidth}
        \includegraphics[width=0.95\textwidth]{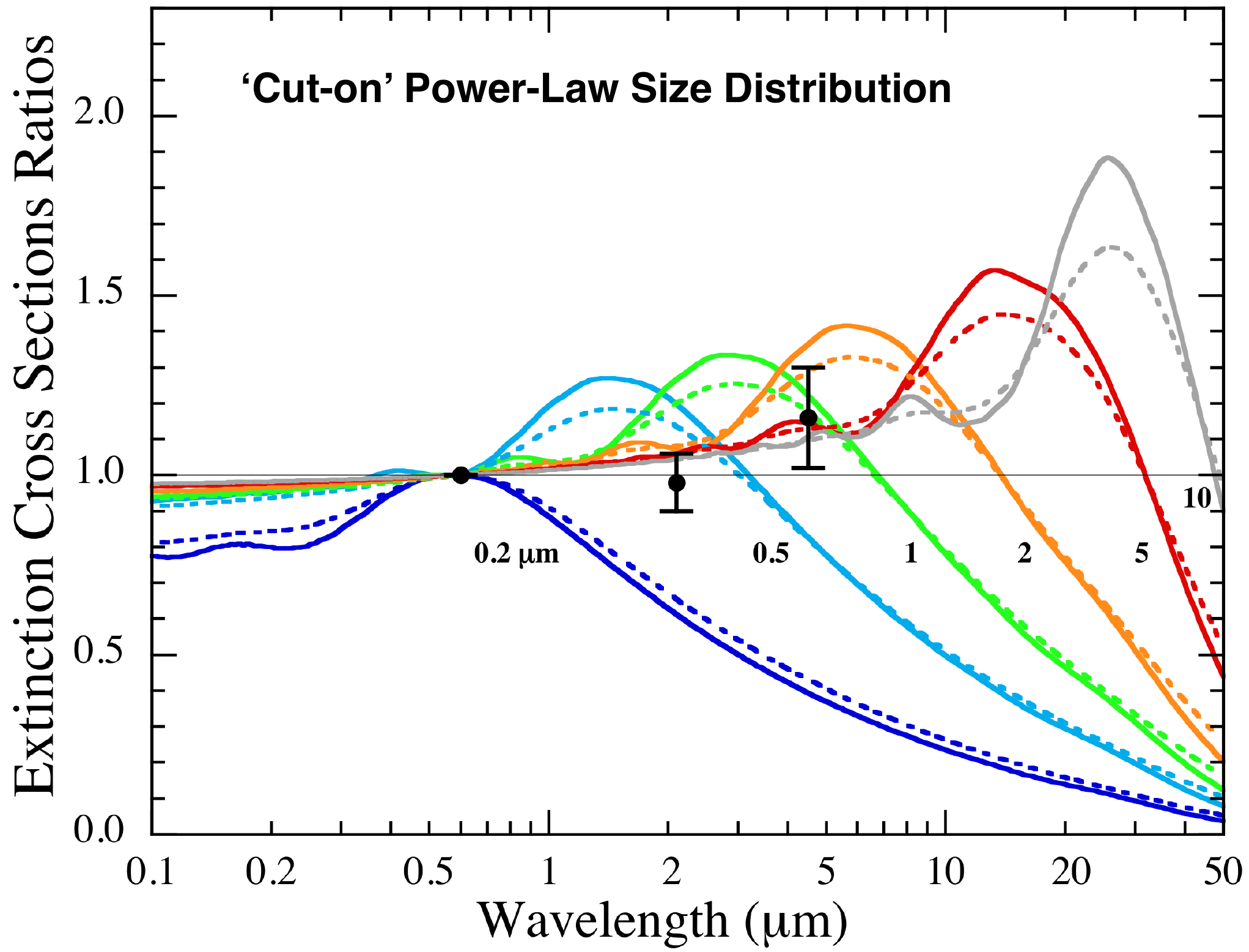} 
    \end{minipage}
\caption{Mie extinction cross sections at wavelength $\lambda$ divided by the corresponding cross section at 0.6~$\mu$m (optical) for a generic grain material, where $n$ is set to be 1.6, and $k=0.1$ (solid curves) and 0.01 (dashed curves). Two size distributions are shown. Left: Hansen distribution, where the labels represent the characteristic particle sizes. Right: a `cut-on' power-law size distribution (see definition in Eq.~\eqref{eq:cuton}), where the labels represent the minimum grain size $s_{min}$. We also show the measured transit depth ratios in WD~1145+017 (black dots) as measured in this work. Only characteristic particle sizes of $\gtrsim 2 ~ \mu$m are consistent with the data. }
 \label{Fig:HansenCuton}
\end{figure*}

As shown in Fig. \ref{Fig:HansenCuton}, we find that for both size distributions, only grains larger than $\gtrsim 2$ $\mu$m can explain the observed transit depth ratios at the different wavelengths.

This minimum inferred grain size, for either choice of size distribution, is broadly consistent with the grain size derived from modeling the 10~$\mu$m silicate emission feature around other dusty white dwarfs \citep{Reach2009,Jura2009a}. On the other hand, small dust ($s$ $\lesssim$ 1~$\mu$m) is prevalent in the interstellar medium \citep[e.g.][]{Mathis1977}, while in the solar system the tails of comets and streams coming off the active asteroids can be a mixture of small and large grains (with $s$ from sub-$\mu$m to cm), depending on the dust production mechanism \citep[e.g.][]{Jewitt2012}.

\section{A Possible Model: Grain Sublimation}
\label{sec:model}

In this section we explore a physical explanation for the
dearth of small grains in the transiting dust clouds.
In summary, the explanation is that small grains have
higher equilibrium temperatures than large grains,
with the result that they are quickly destroyed by sublimation (see also \citealt{vonHippel2007}).
The process of sublimation has an extremely steep dependence on temperature,
so a modest increase in dust temperature can result in drastically shorter sublimation timescales.
To assess this scenario, we first compute the equilibrium temperature
of dust grains in the transiting clouds (Sect.~\ref{sec:temp_dust}),
then evaluate whether sublimation occurs in the potentially gas-rich
circumstellar environment of WD~1145+017 (Sect.~\ref{sec:subl_bal}),
and finally calculate dust sublimation timescales as a function of grain size (Sect.~\ref{sec:t_subl}).
Throughout this analysis we use the parameter values for the WD 1145+017 system listed in Table~\ref{Tab:WDParameters}.

\subsection{Dust Temperatures}
\label{sec:temp_dust}

The temperature $ T_\mathrm{d}( s, r ) $ of a dust grain with radius $ s $ at distance $ r $ from the white dwarf can be calculated by solving the power balance between the incoming stellar radiation and the outgoing thermal radiation:
\begin{equation}
  \label{eq:temp_d}
  \frac{ R_\star^2 }{ 4 r^2 } \int Q_\mathrm{abs}( s, \lambda ) B_\lambda( \lambda, T_\star ) \, \mathrm{d} \lambda
    = \int Q_\mathrm{abs}( s, \lambda ) B_\lambda( \lambda, T_\mathrm{d} ) \, \mathrm{d} \lambda.
\end{equation}
The meanings for some symbols are listed in Table \ref{Tab:WDParameters}. $ Q_\mathrm{abs} $ is the absorption efficiency of the dust grain, and $ B_\lambda $ is the Planck function. The white dwarf spectrum is approximated by blackbody radiation. This calculation ignores the latent heat of sublimation, which we find to be negligible \citep{Rappaport2014}.

The grain temperature depends critically on the absorption efficiency $ Q_\mathrm{abs} $,
which is generally a complicated function of grain size $s$, wavelength $\lambda$,
and the optical constants of the dust material, i.e., its complex refractive index $ n + i k $. In certain cases, however,  simple prescriptions for $ Q_\mathrm{abs} $ can be derived,
which allow the power balance Eq.~\eqref{eq:temp_d} to be solved analytically \citep[e.g.][]{BackmanParesce1993,vonHippel2007}.

For grains that are very large compared to the radiation wavelength,
the absorption efficiency asymptotically approaches
a constant value of $ Q_\mathrm{abs} \approx 1 $.
In this limit, solving Eq.~\eqref{eq:temp_d} gives the blackbody temperature
\begin{align}
  \label{eq:temp_bb}
  T_\mathrm{bb}
    & = \sqrt{ \frac{ R_\star }{ 2 r } } T_\star \\
    & \approx 1.2 \times 10^3 \, \mathrm{K} \;
      \Biggl( \frac{ T_\star }{ \mathrm{15{,}900\,K} } \Biggr) \,
      \Biggl( \frac{ R_\star }{ \mathrm{0.013\,R_\odot} } \Biggr)^{1/2}
      \Biggl( \frac{ r }{ \mathrm{1.16\,R_\odot} } \Biggr)^{-1/2}. \nonumber
\end{align}

For very small grains, which fall in the Rayleigh regime,
$ Q_\mathrm{abs} \propto \lambda^{-1} $ is found,
which yields (e.g., Appendix~C of \citealt{Rappaport2014})
\begin{align}
  \label{eq:temp_rayl}
  T_\mathrm{Rayl}
    & = \left( \frac{ R_\star }{ 2 r } \right)^{2/5} T_\star \\
    & \approx 2.0 \times 10^3 \, \mathrm{K} \;
      \Biggl( \frac{ T_\star }{ \mathrm{15{,}900\,K} } \Biggr) \,
      \Biggl( \frac{ R_\star }{ \mathrm{0.013\,R_\odot} } \Biggr)^{2/5}
      \Biggl( \frac{ r }{ \mathrm{1.16\,R_\odot} } \Biggr)^{-2/5}. \nonumber
\end{align}

\begin{figure}
 \includegraphics[width=\linewidth]{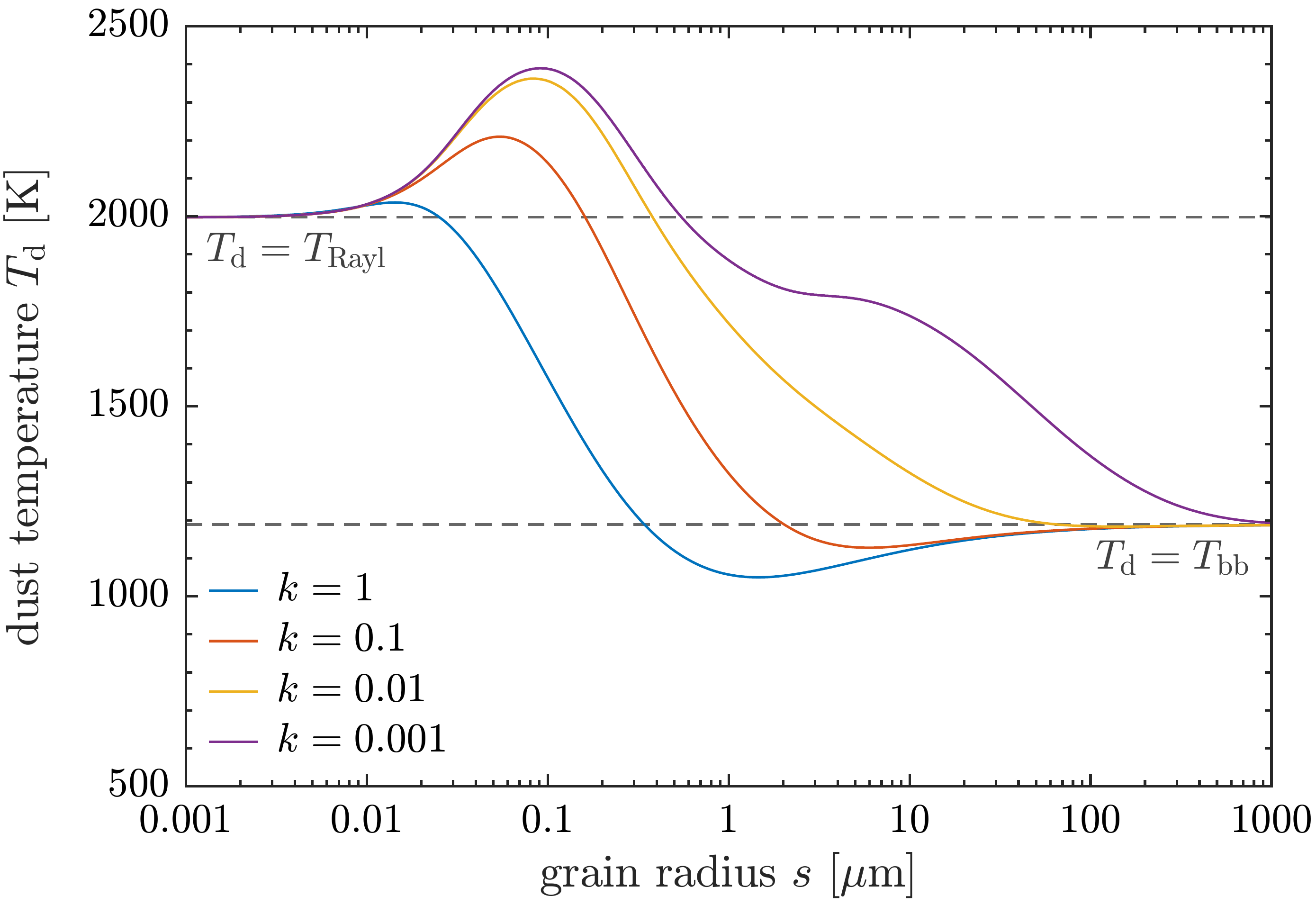}
  \caption{Temperature as a function of grain size for dust grains in orbit around WD~1145+017 at 4.5 hr. The solid coloured lines represent generic minerals with different values of $k$, the imaginary part of the index of refraction.  The real part is kept fixed at n = 1.6.  Also shown are the two limiting cases of Rayleigh-approximation and blackbody temperatures (dashed horizontal lines). Small grains are heated to a higher temperature than large grains.}
  \label{Fig:size_temp}
\end{figure}

For general cases, the grain temperature can be found by solving Eq.~\eqref{eq:temp_d} using $ Q_\mathrm{abs} $ from the Mie theory \citep{BohrenHuffman1983}. In Fig.~\ref{Fig:size_temp} we plot grain temperatures as a function of grain size. This shows the convergence towards the limiting temperatures for small and large grain sizes. Small grains can reach a much higher temperature of 2000~K than large grains of 1200~K. Interestingly, for dust orbiting WD 1145+017 in a 4.5 hr period, the values of the two limiting temperatures bracket the temperatures at which many refractory materials sublimate rapidly.

\subsection{The Sublimation/Condensation Balance}\label{sec:subl_bal}

Like all thermodynamic phase transitions,
sublimation depends on the pressure and temperature of the matter involved,
and the balance between sublimation and condensation can be evaluated using a phase diagram.
For a given temperature $ T $, the pressure at which these two processes are in equilibrium
is called the phase-equilibrium (or saturated) vapour pressure $ p_\mathrm{sat} $.
Based on the Clausius--Clapeyron relation, its temperature dependence is found to be
\begin{equation}
  \label{eq:pres_vap}
  p_\mathrm{sat}( T ) = \exp ( \, - \mathcal{A} / T + \mathcal{B} \, ),
\end{equation}
where $ \mathcal{A} $ and $ \mathcal{B} $ are material-dependent sublimation parameters
that can be determined experimentally.
Dust grains with temperature $ T_\mathrm{d} $ will sublimate when the ambient gas pressure
$ p_\mathrm{g} $ is lower than $ p_\mathrm{sat}( T_\mathrm{d} ) $,
while condensation happens for $ p_\mathrm{g} > p_\mathrm{sat}( T_\mathrm{d} ) $.

To assess whether sublimation or condensation dominates,
we make a rough estimate of the ambient gas pressure.
Assuming the gas around the white dwarf forms a steady-state viscous accretion disc,
where the viscosity is parameterised by $ \alpha_\nu $,
the gas pressure in the disc is given approximately by \citep{RafikovGarmilla2012}
\begin{align}
  \label{eq:pres_g}
  p_\mathrm{g}
    & = \frac{ \dot{M} \varOmega_\mathrm{K}^2 }{ 3 \uppi \alpha_\nu c_\mathrm{s} } \\
    & \approx 0.2 \, \mathrm{dyn\,cm^{-2}} \;
      \Biggl( \frac{ \dot{M} }{ \mathrm{10^{10}\,g\,s^{-1}} } \Biggr) \,
      \Biggl( \frac{ P }{ \mathrm{4.5\,hr} } \Biggr)^{-2}
      \Biggl( \frac{ \alpha_\nu }{ 0.01 } \Biggr)^{-1}
      \Biggl( \frac{ c_\mathrm{s} }{ \mathrm{1~km~s^{-1}} } \Biggr)^{-1}. \nonumber
\end{align}
Here, $ \varOmega_\mathrm{K} $ is the local Keplerian angular frequency
and $ c_\mathrm{s} $ is the sound speed.
The mass accretion rate $ \dot{M} $ can be estimated from pollution in the white dwarf's atmosphere, assuming a steady state accretion \citep{Xu2016}. In addition, $ c_\mathrm{s} \sim \mathrm{1~km\,s^{-1}} $ is a reasonable estimate
for the sound speed of a metallic gas with a temperature of a few thousand K,
so the uncertainty in $ p_\mathrm{g} $ is dominated by the poor knowledge of $ \alpha_\nu $.  Note that the temperature and vertical distribution of the gaseous disc can be substantially different from that of the dust (sect.~6 in \citealt{Melis2010}). 

\begin{figure}
  \centering
  \includegraphics[width=\linewidth]{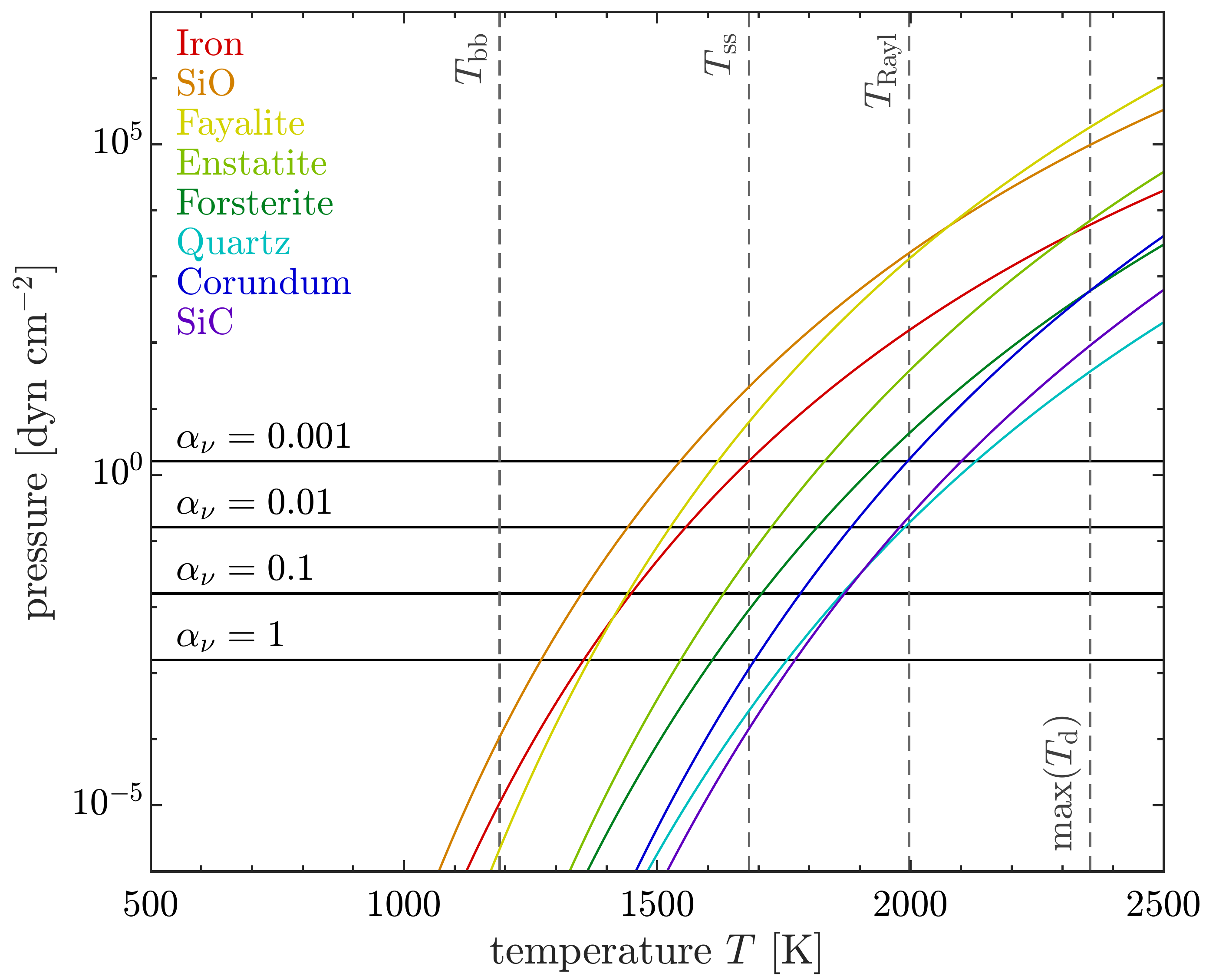}
  \caption{Phase-equilibrium vapour pressure $ p_\mathrm{sat} $ as function of temperature
  for a set of possible dust species from Eq. \eqref{eq:pres_vap} (coloured lines),
  together with estimates of the ambient gas pressure $ p_\mathrm{g} $ for several different
  values of the viscosity parameter $ \alpha_\nu $ from Eq. \eqref{eq:pres_g} (horizontal black lines). Sublimation occurs when $ p_\mathrm{g} $ < $ p_\mathrm{sat}$, while condensation happens when $ p_\mathrm{g} $ > $ p_\mathrm{sat}$.
  Also indicated are temperatures reached by particles orbiting \mbox{WD~1145+017}
  in several limiting cases (dashed vertical lines):
  $ T_\mathrm{bb} $, the blackbody temperature valid for large dust grains;
  $ T_\mathrm{ss} $, the temperature at the substellar point of a tidally locked large body (based on eq.~(5) in \citealp{Vanderburg2015});
 $ T_\mathrm{Rayl} $, the limiting temperature for small grains;
  and $ \max( T_\mathrm{d} ) $, the maximum dust temperature
  seen in Fig.~\ref{Fig:size_temp}.}
  \label{Fig:pres_temp}
\end{figure}

Fig.~\ref{Fig:pres_temp} shows a comparison of the phase-equilibrium vapour pressures of a set of possible refractory materials, excluding graphite\footnote{Graphite is unlikely to be the dominant component of the dust particles because carbon has not yet been detected in the atmosphere of WD 1145+017 \citep{Xu2016}. In fact, almost all polluted white dwarfs are carbon-depleted \citep[e.g.][]{Jura2006}. Graphite is excluded for all the following analyses.}(using values of $ \mathcal{A} $ and $ \mathcal{B} $ in Table 3 of \citealt{vanLieshout2014}) with the estimated ambient gas pressure associated with the presumed gaseous accretion disc. Although there is great uncertainty in both $ p_\mathrm{sat} $ (because the dust composition is not well constrained) and $ p_\mathrm{g} $ (because $ \alpha_\nu $ is unknown), the figure reveals that sublimation is expected for dust with temperatures around $ T_\mathrm{Rayl} $ (small grains), while material with temperatures around $ T_\mathrm{bb} $ (large grains) could be protected from sublimation by the gas disc.

\subsection{A Minimum Grain Size Due to Sublimation}
\label{sec:t_subl}

Because the Rayleigh-approximation temperature is higher than the blackbody temperature
($ T_\mathrm{Rayl} / T_\mathrm{bb} \approx 1.7$),
the dust temperature must go up with decreasing grain size.
Dust sublimation rates have an extremely steep dependence on temperature,
so this increase in temperature will be associated with
a dramatic decrease in dust survival time against sublimation.
In contrast, when the dust temperature is constant with grain size,
the sublimation timescale will only decrease linearly with decreasing grain size.

To quantify the effect of sublimation on grain survival times,
we compute dust sublimation timescales, given by
\begin{equation}
  \label{eq:t_subl}
  t_\mathrm{subl}
    = - \frac{ s }{ \dot{s} }
    = \frac{ s \rho_\mathrm{d} }{ J( T_\mathrm{d} ) }.
\end{equation}
Here,
$ s $ is the grain radius,
$ \dot{ s } $ is its change rate,
$ \rho_\mathrm{d} $ is the density of the dust material,
and $ J $ is the net sublimation mass-loss flux
(units: [g~cm$^{-2}$~s$^{-1}$]; positive for mass loss).
The mass-loss flux $ J $ can be calculated from the kinetic theory of gasses \citep[e.g.][]{Langmuir1913}:
\begin{equation}
  \label{eq:j_net}
  J( T ) = \alpha_\mathrm{subl} \left[ p_\mathrm{sat}( T ) - p_\mathrm{g} \right]
      \sqrt{ \frac{ \mu m_\mathrm{u} }{ 2 \pi k_\mathrm{B} T } }.
\end{equation}
Here,
$ \alpha_\mathrm{subl} $ is the evaporation coefficient, also known as the `accommodation coefficient' or `sticking efficiency',
which parametrises kinetic inhibition of the sublimation process (which we assume to be independent of temperature),
$ \mu $ is the molecular weight of the molecules that sublimate,
$ m_\mathrm{u} $ is the atomic mass unit,
and $ k_\mathrm{B} $ is the Boltzmann constant.

\begin{figure}
  \centering
  \includegraphics[width=\linewidth]{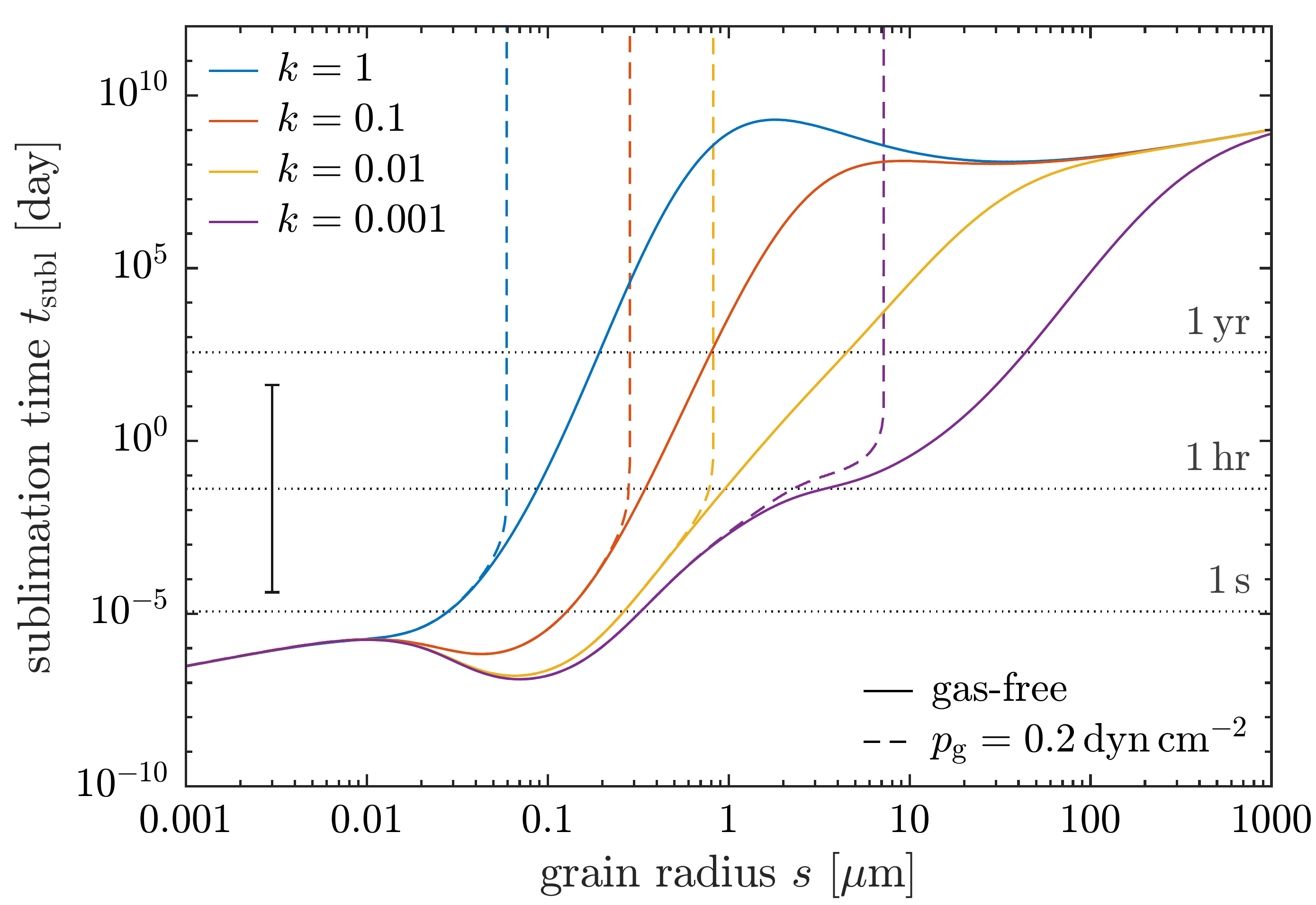}
  \caption{Sublimation timescale as function of grain size
  assuming the dust temperatures from Fig.~\ref{Fig:size_temp}
  and sublimation parameters corresponding to a generic refractory material.
  Different coloured lines correspond to different values of
  the imaginary part of the complex refractive index $ k $.
  The real part is kept fixed at $ n = 1.6 $.
  The solid curves are for dust grains in vacuum;
  the dashed curves assume an ambient gas density of
  $ p_\mathrm{g} \approx 0.2 \mathrm{\,dyn\,cm^{-2}} $ (see Sect.~\ref{sec:subl_bal}).
  The error bar is a rough indication of the uncertainty in sublimation time
  found by considering different possible dust materials
  (specifically, those listed in Fig.~\ref{Fig:pres_temp}.)}
  \label{Fig:size_tsubl}
\end{figure}

In Fig.~\ref{Fig:size_tsubl}, we show the sublimation timescale as a function
of grain size using dust temperatures computed in Sect.~\ref{sec:temp_dust}.
The calculation uses a set of sublimation parameters
typical for a generic refractory material:
$ \mathcal{A} = 65{,}000 $\,K,
$ \mathcal{B} = 35 $,
$ \rho_\mathrm{d} = 3 \mathrm{\,g\,cm^{-3}} $,
$ \alpha_\mathrm{subl} = 0.1 $,
and $ \mu = 100 $.
These fall roughly in the middle of the range of values seen for
the refractory materials shown in Fig.~\ref{Fig:pres_temp}
(see \citealt{vanLieshout2014}). 

Fig.~\ref{Fig:size_tsubl} demonstrates that small dust is destroyed almost instantaneously,
while large dust could survive against sublimation for many years.
This result
is robust
despite the large uncertainty in sublimation timescale
introduced by the uncertainty in dust composition and hence sublimation parameters
(there are about 2 to 4 orders of magnitude spread in sublimation timescale
amongst the materials shown in Fig.~\ref{Fig:pres_temp}).
We conclude that the large grain sizes inferred from
the lack of wavelength dependence in the transit depths of WD 1145+017
is likely the result of sublimation of smaller grains,
because of their higher equilibrium temperatures and rapid sublimation.
This conclusion holds when considering both gas-free and gas-rich environments.
In the gas-free case,
large grains have very long sublimation timescales,
and their lifetime is likely set by other destruction processes than sublimation.
In the gas-rich case,
large grains are also protected from sublimation by the gas disc (as discussed in Sect.~\ref{sec:subl_bal}).

Our method of estimating sublimation timescales assumes the temperature
of the dust grain to remain constant as it decreases in size due to sublimation,
which is incorrect as shown in Fig.~\ref{Fig:size_temp}.
However, given the extreme temperature dependence of the sublimation process,
the sublimation timescale will be dominated by the lowest dust temperatures encountered.
Hence, Eq.~\eqref{eq:t_subl} using the initial dust temperature
provides a very good estimate of the sublimation timescale.

The exact value of the grain size $ s $ below which sublimation
destroys grains faster than they are replenished depends on a number of factors:
the optical properties of the dust
(most importantly the value of the imaginary part of the complex refractive index $ k $);
its sublimation parameters (i.e. $ \mathcal{A} $, $ \mathcal{B} $, $\alpha_\mathrm{subl}$, $\mu$, and $\rho_\mathrm{d}$);
the timescale on which other processes (like collisions) destroy dust grains
when the sublimation timescale is long;
and the size-dependent input rate of dust, which is determined by the dust production process and is still unknown.
By modelling the resultant grain size distribution, it is in principle possible
to use the lower limit on the minimum grain size inferred from the observation
to put constraints on the composition of the dust.
This exercise is beyond the scope of the present work,
but will be the subject of a future study.
For now, we tentatively suggest that the inferred lower limit of
$ s $  $\gtrsim$ 1.5 $\mu$m
disfavors metallic dust species like pure iron\footnote{In spite of this conclusion, we note that Fe is abundant in the atmosphere of WD~1145+017 \citep{Xu2016}.}.
The reason is that these materials typically have $ k$  > 1 (i.e., they are reflective) and grains smaller than 1 $\mu$m can survive for a considerable amount of time before sublimation, as shown in Fig.~\ref{Fig:size_tsubl}.

\section{Infrared Excess and the Transiting Material}
\label{sec:IRExcess}

\begin{figure}
\includegraphics[width=\linewidth]{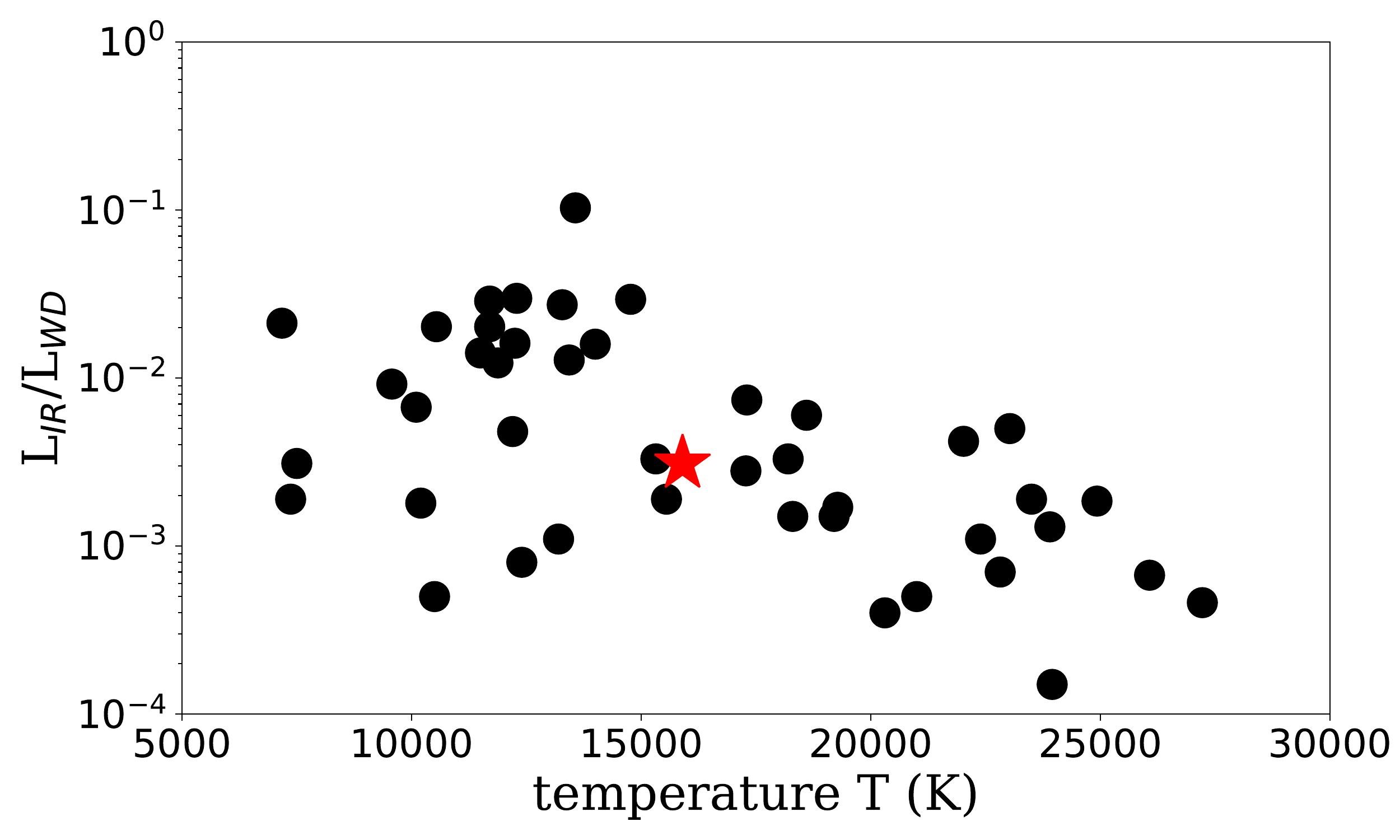}
\caption{Fractional luminosity of the infrared excess for all known dusty white dwarfs versus the effective temperature. The red star represents WD~1145+017, whose fractional IR luminosity is typical of dusty white dwarfs at this temperature range.}
\label{Fig:TauT}
\end{figure}

WD~1145+017 displays excess infrared radiation starting from the K$_\mathrm{s}$ band. There are over 40 known dusty white dwarfs \citep{Farihi2016} and usually, they have been modelled with a geometrically thin, optically thick dust disc within the white dwarf's tidal radius \citep{Jura2003}. The sublimation/condensation calculation presented in Sect.~\ref{sec:model} can also be relevant for the innermost region of the dust discs, where the dust is optically thin and directly exposed to the radiation from the white dwarf. However, the detailed process is highly dependent on the density and viscosity of the surrounding metallic gas \citep[e.g.][]{Rafikov2011b}. Occasionally, the infrared excess is so strong that a warped disc is preferred \citep{Jura2007a,Jura2007b}. As shown in Fig. \ref{Fig:TauT}, the fractional luminosity of the infrared excess around WD~1145+017 is comparable to other white dwarfs with a dust disc.  There is a general trend of increasing fractional luminosity as the white dwarf cools, which points to a possible disc evolution sequence \citep{Rocchetto2015}.

The {\it Spitzer} IRAC-2 measurement of 55.0~$\pm$~3.2~$\mu$Jy at 4.5 $\mu$m greatly reduced the uncertainty compared to the WISE-2 measurement of 43~$\pm$~14~$\mu$Jy (see Sect.~\ref{sec: LC_Spitzer}). Here, we explore the spectral energy distribution (SED) fits of WD~1145+017 with two simple models.

\begin{figure}
\includegraphics[width=\linewidth]{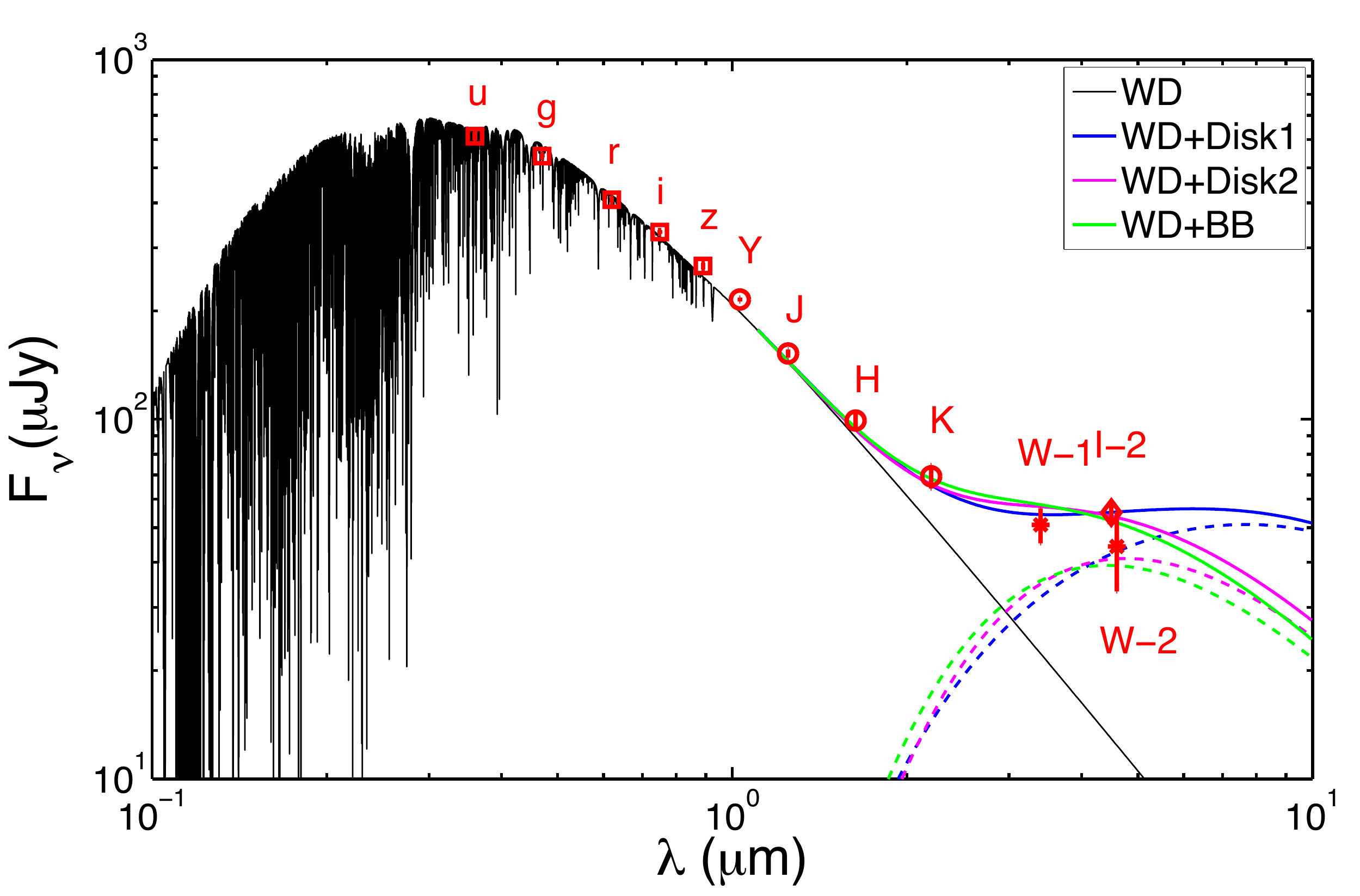}
\caption{SED fits of the infrared excess around WD 1145+017. The red square, circle, star, and diamond symbols represent measurement from the SDSS, UKIDSS, WISE, and {\it Spitzer}, respectively. The black line represents flux from the white dwarf model. The blue and magenta dashed lines represent flux from an opaque disc with parameters listed in Table \ref{Tab:discFit} while the blue dashed line represents flux from a blackbody. The solid coloured lines represent flux from the white dwarf plus an additional component, either a disc or a blackbody. }
\label{Fig:SED}
\end{figure}

{\it (i) A flat opaque disc.} We follow the classical recipe of fitting opaque dust discs around white dwarfs from \citet{Jura2003} with three free parameters, the inner radius of the disc $R_\mathrm{in}$, outer radius of the disc $R_\mathrm{out}$, and inclination $\mathrm{cos}\ i$. There is a degeneracy between the surface area of the dust disc and its inclination. We performed chi-squared minimizations to find the best-fit parameters for two extreme cases, i.e. with large and small inclinations, as listed in Table \ref{Tab:discFit} and also shown in Fig. \ref{Fig:SED}. For model ``disc1'', the inner radius corresponds to a temperature similar to the inner disc temperature of other dusty white dwarfs, likely determined by sublimation and the outer radius is located near the tidal radius. Model ``disc2'' is effectively a face-on narrow ring close to the white dwarf. Its SED is very similar to that of a blackbody. To produce the observed infrared excess with a flat opaque disc, the disc would not be aligned with the transiting objects. Under this scenario, either the dust disc and the transiting objects come from different parent bodies with different orbital inclination, or some additional mechanism is required to perturb the dust disc to be misaligned with the transiting objects.

{\it (ii) An inflated optically thin disc.} The best-fit blackbody model has an effective temperature of 1150 K and surface area of 160$\pi R_\star^2$, which is consistent with the numbers derived in \citet{Vanderburg2015}. This temperature is comparable to the temperature of large grains around the transiting material at 1.16R$_{\sun}$ (90\,$R_{\star}$), as derived in Eq.~\eqref{eq:temp_bb}. To produce the observed infrared excess, a disc height of 0.9\,$R_\star$ is required. Collisional cascades around the Roche limits of white dwarfs have been studied recently in \citet{KenyonBromley2017}. They found for discs made of indestructible particles, the scale height would quickly reduce to a value that is comparable to the particle radius. However, with additional mass input, the scale height of the dust disc can remain quite high for a long time. This is a viable alternative because there is a constant mass input from the disintegrating material into the dust disc around WD~1145+017. In addition, from the deep transits in the light curve, we know that the transiting material has significant height as well. A prediction from this model is that the disc scale height is dependent on the mass input rate. It is essential to keep monitoring the infrared flux of the dust disc to look for any variations correlated with the transit light curve.

\begin{table}
  \centering
  \caption{Best-fit parameters to the SED of WD~1145+017}
  \label{Tab:discFit}
  \begin{tabular}{lclc}
  \hline
  \hline
  Model & $\chi^2_\mathrm{d}$ & Parameters \\
  \hline
  disc1 & 1.3	& R$_\mathrm{in}$=13R$_\star$, R$_\mathrm{out}$=120R$_\star$, cos $i$ = 0.18\\
  	&	& T$_\mathrm{in}$=1570 K, T$_\mathrm{out}$=300 K \\
  disc2 &	2.5	& R$_\mathrm{in}$=19R$_\star$, R$_\mathrm{out}$=25R$_\star$, cos $i$=0.80\\
	&	& T$_\mathrm{in}$=1190 K, T$_\mathrm{out}$=970 K\\
  Blackbody	& 1.3		& T$_\mathrm{bb}$=1150 K, A=160$\pi R_\star^2$\\
  \hline
  \end{tabular}
  \\
  {\bf Note.} \\
 We fit four data points, including fluxes from H, K$_\mathrm{s}$, W-1, and IRAC-2. $\chi^2_\mathrm{d}$ is calculated for per degree of freedom, which is 1 for the opaque disc model and 2 for the blackbody model. For disc1, the outer radius R$_\mathrm{out}$ is not well constrained due to the lack of longer wavelength observations. We kept it at 120R$_\star$, the typical tidal radius of white dwarfs. 
\end{table}

It is worth noting that the transiting material could contribute to the infrared excess as well. We approximate the total effective surface area of the dust as a cylinder and it can be calculated as:
\begin{equation}
  \label{eq:area_dust}
  A   \approx 2 \pi r \cdot \epsilon \cdot \delta \cdot \sqrt{\pi} R_\star
\end{equation}
where $\epsilon$ is the percentage of time a transit lasts, $\delta$ is the average transit depth, both can be estimated from the light curve and $\sqrt{\pi} R_\star$ represents the effective height of the stellar disc. For the 2016 dataset, we found $\epsilon$ = 0.22 and $\delta$= 0.17. For the 2017 dataset, for B2, $\epsilon$ = 0.10, $\delta$ = 0.09 while for B3, $\epsilon$ = 0.25, $\delta$=0.33. The total surface area of the transiting material is $\sim$~12$\pi R_\star^2$ and $\sim$~29$\pi R_\star^2$ in 2016 and 2017, respectively. Thus, there is an increase in surface area by a factor of 2.5 in the 2017 light curve compared to the 2016 light curve. However, the surface area of the transiting material is still relatively small compared to the total inferred surface area of the emitting dust of 160$\pi R_\star^2$, and the 4.5 $\mu$m flux remains constant between 2016 and 2017 observations, requiring a long lasting reservoir of $\sim$1200\,K dust. Regardless, there could be a significant fraction of non-transiting dust material and the sublimation timescale for large particle is years or longer, as shown in Fig. \ref{Fig:size_tsubl}. The transiting objects could supply material for the observed infrared excess.

\section{Summary \& Conclusion}
\label{sec:conclusion}

We have presented two epochs of multi-wavelength photometric observations of WD~1145+017, covering from optical to 4.5~$\mu$m during 2016 March 28-29 and 2017 April 4-5. One main transit feature was detected in 2016 and three transit features in 2017 during a 4.5~hr orbital period. We modelled the transit features and found that the observed transit depths were different at all the observed wavelengths. After correcting for the excess infrared emission from orbiting dust particles in the K$_\mathrm{s}$ and 4.5~$\mu$m bands, we found that the transit depths are the same from optical to 4.5~$\mu$m during both epochs. 

This wavelength-independent transit behaviour can be explained by a dearth of small grains ($\lesssim$ 1.5~$\mu$m) in the transiting material. Small grains are heated to a higher temperature than large grains. In addition, sublimation rates have a steep dependence on grain temperature. As a result, we suggest a model where small grains sublimate rapidly and only large grains can survive long enough to be detectable. 

The dust released from the hypothesised orbiting bodies will continually add mass into the circumstellar dust, which could potentially maintain a large scale height. We present two models that can equally fit the infrared excess of WD~1145+017, including a flat opaque disc and an inflated optically thin disc. Future observations, particularly monitoring in the infrared, will improve our understanding of the link between the transiting material and the infrared excess.

\vspace{0.2cm}
\noindent 
{\bf Note Added in Manuscript}: After this work was substantially complete, we became aware of an interesting model by \citet{Farihi2017} that might explain some of the properties of the peculiar transits in WD~1145+017. This model involves magnetic entrainment of small ($a$ $\sim$~0.1~$\mu$m) charged particles in the field of a strongly magnetised white dwarf. In the context of either the \citet{Farihi2017} scenario, or the one we have assumed, namely dust clouds released by orbiting bodies with periods near 4.5 hours, all of the measured transit depths and dust size determinations made in this work should still be equally valid. All of the calculations we made suggesting that the smaller dust grains should sublimate quickly would remain unchanged in either scenario.  Thus, none of our conclusions should be affected.

\section*{Acknowledgements}
We thank the {\it Spitzer} helpdesk for useful discussions about high-precision photometry. The paper was based on observations made with: (i) the {\it Spitzer Space Telescope} under program \#12128 and \#13065, which is operated by the Jet Propulsion Laboratory, California Institute of Technology under a contract with NASA. Support for this work was provided by NASA through an award issued by JPL/Caltech. (ii) the European Organisation for Astronomical Research in the Southern Hemisphere under ESO programs 296.C-5024 and 099.C-0082. This work also makes use of observations from the LCO network. R. van Lieshout acknowledges support from the European Union through ERC grant numbers 279973 and 341137. A. Camero acknowledges support from STFC grant ST/M001296/1.


\bibliographystyle{mnras}




\appendix

\section{Optical Light Curve Comparison}
\label{sec:app}

\begin{figure}
\includegraphics[width=\linewidth]{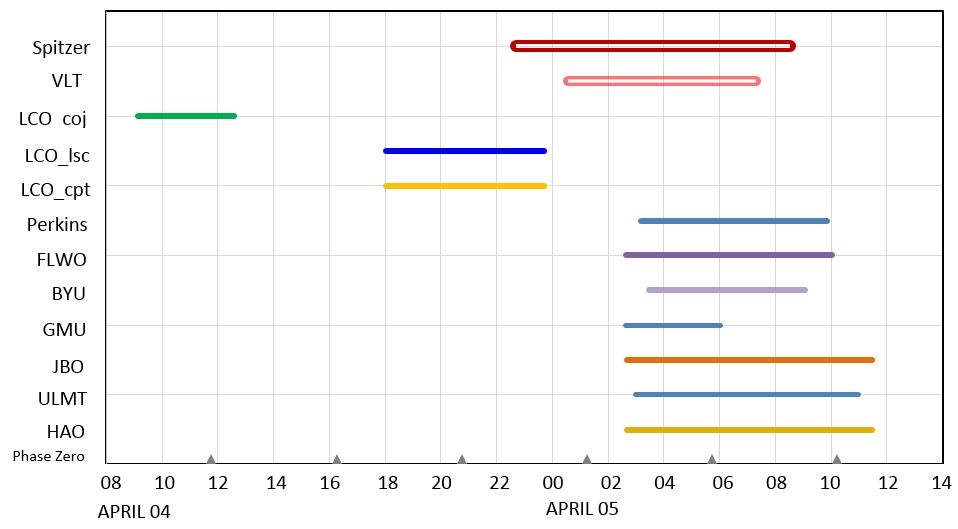}
\caption{Observing times for 10 optical sessions in 2017. The details of the observation are listed in Table \ref{Tab:OpticalObs2017}. The {\it Spitzer} and {\it VLT} observing windows are also shown for comparison. The triangles indicate phase zero, as defined in the text.
}
\label{Fig: LC_Optical_Comp}
\end{figure}

In support of the {\it Spitzer} observation on April 4-5, 2017, we requested observing time in ten optical telescopes all around the world. These 10 ground-based light curves offer a unique opportunity for assessing the level of systematics that can be expected in any single light curve. The observations are summarised in Table \ref{Tab:OpticalObs2017}. These 10 observing sessions spanned 26 hours, or $\sim$ 6 WD 1145+017 orbital periods, as shown in Fig. \ref{Fig: LC_Optical_Comp}.

\begin{figure}
\includegraphics[width=\linewidth]{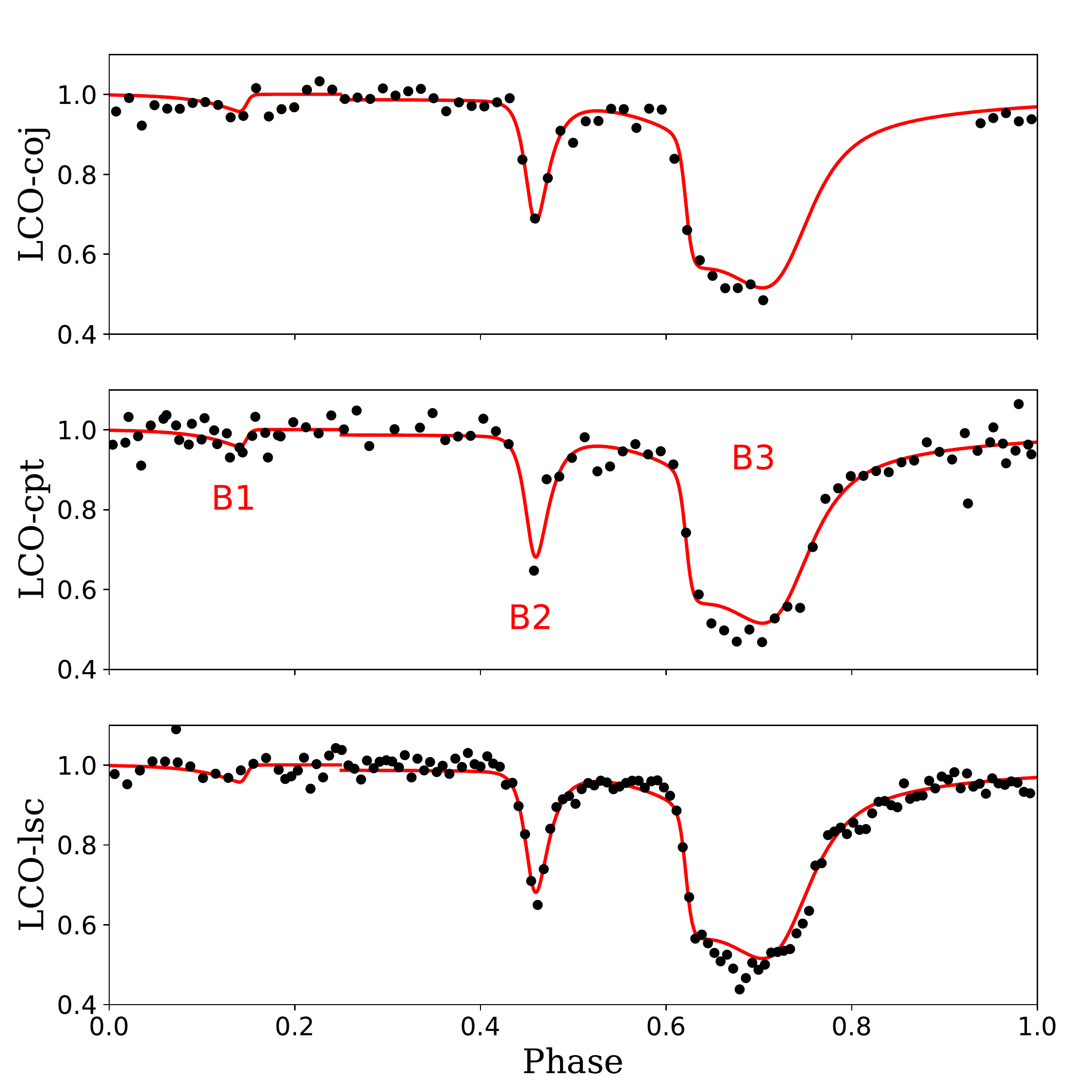}
\caption{Comparison of the 3 light curves in the early group. The red line represents the best-fit model for the Perkins data, as described in Sect.~\ref{sec: FitDepth}. All three light curves have a deeper B3 dip than the Perkins model, which is likely due to evolution of the transit shape over time.
}
\label{Fig: Optical_EG}
\end{figure}

\begin{figure}
\includegraphics[width=\linewidth]{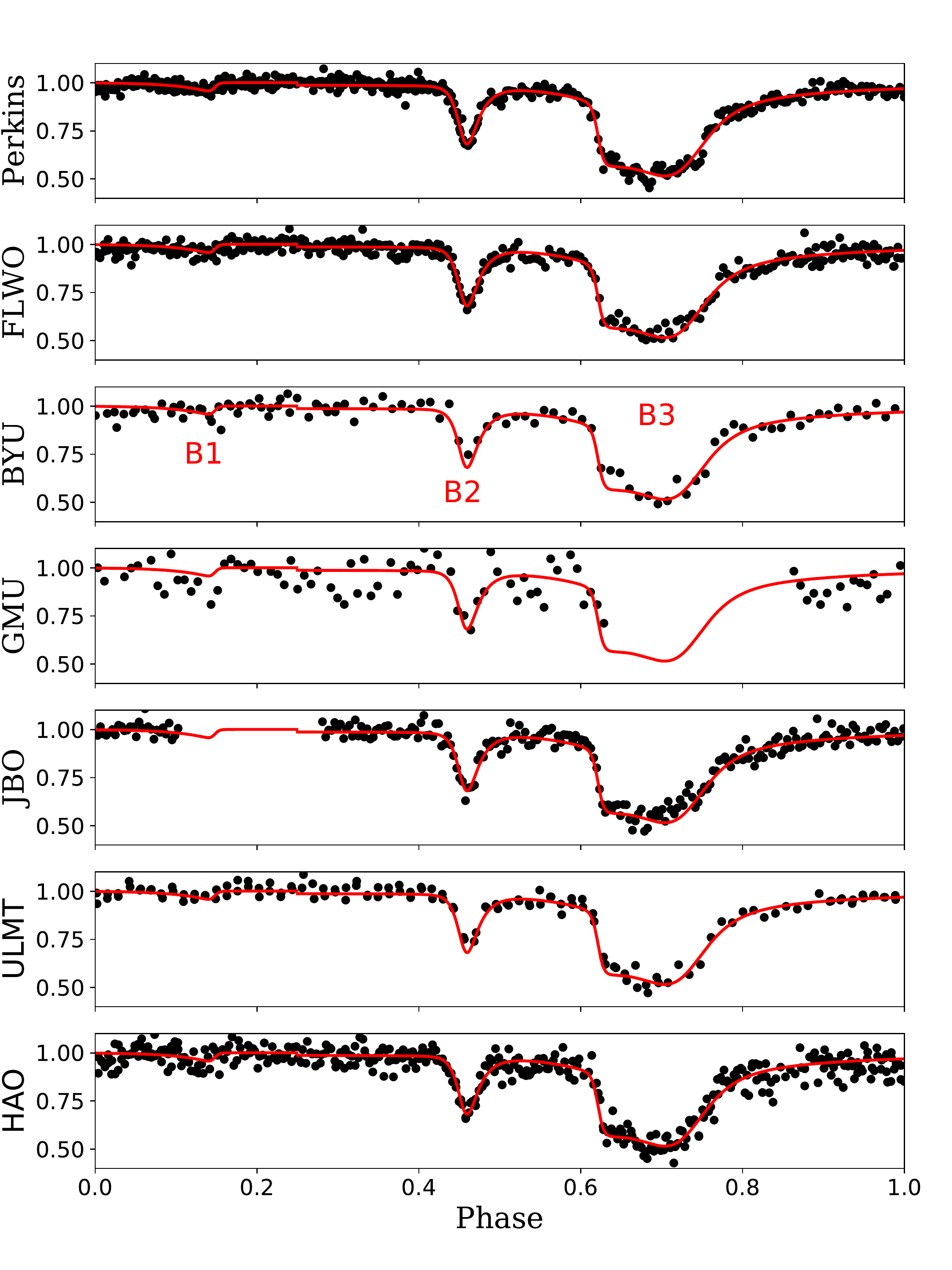}
\caption{Comparison of the 7 light curves in the late group. The red line represents the best-fit model for the {\it Perkins} data, as described in Sect.~\ref{sec: FitDepth}.
}
\label{Fig: Optical_LG}
\end{figure}

\begin{table*}
\centering
\caption{Summary of Optical Observations on April 4-5, 2017}
\label{Tab:OpticalObs2017}
\begin{tabular}{llllllll}
\hline
\textbf{Telescope Name} & \textbf{Location}             & \textbf{Aperture} & \textbf{Observers}               & \textbf{Cadence} & \textbf{Filters}   \\ \hline
\multicolumn{6}{c}{\bf Early Group }\\
LCO/COJ               & SSO, Australia            & 1m                &Avi Shporer & 221s                       & g                                      \\
LCO/LSC                 & CTIO, Chile           & 1m                & Avi Shporer & 219s                       & g                                  \\
LCO/CPT                 & SAAO, South Africa            & 1m                & Avi Shporer & 220s                       & g                                   \\
\\
\multicolumn{6}{c}{\bf Late Group} \\
Perkins                 & Lowell, AZ                    & 1.8m              & Paul Dalba                       & 51s                        & R                       \\
FLWO                    & FLWO, AZ                      & 1.2m              & Allyson Bieryla                  & 77s                        & V                                \\
BYU                     & West Mountain Observatory, UT & 0.91m             & Michael Joner                    & 190s                       & V                                 \\
GMU                     & Fairfax, VA                   & 0.81m             & Jenna Cann, Peter A. Panka       & 133s                       & Clear                               \\
JBO                     & Hereford, AZ                  & 0.81m             & Tom Kaye                         & 58s                        & Clear                            \\
ULMT                    & Mt. Lemmon, AZ                & 0.6m              & Karen Collins                    & 213s                       & Clear             \\             
HAO                     & Hereford, AZ                  & 0.35m             & Bruce Gary                       & 87s                        & Clear                              \\
\hline
\end{tabular}
\end{table*}
Most of the data were reduced with \textsc{AstroImageJ}, an open software for high-precision light curve extrapolations \citep{CollinsKielkopf2013, Collins2017}. The Las Cumbres Observatory (LCO) data were reduced following previous procedures developed for LCO observations of WD~1145+017 outlined in \citet{Zhou2016}. The data reduction procedures for the JBO and HAO observations were described in \citet{Rappaport2016}. The Perkins observations were described in detail in Sect.~\ref{sec:OpticalObs}. All the light curves are shown in Figs \ref{Fig: Optical_EG} and \ref{Fig: Optical_LG}, also shown is the best-fit model for the {\it Perkins} data.

There are three main dip features in an orbital period, which we name B1, B2, and B3, respectively. Since the transits around WD~1145+017 change gradually, any noticeable change typically occurs on a longer timescale than an orbital period. We consider the first 3 light curves as close enough in time to justify a direct comparison. The last 7 light curves are simultaneous. We will refer to those two groups as `early group' and `late group'. 

The early group of light curves were all obtained with the 1-m LCO network \citep{Brown2013}. The variations among the light curves suggest that systematic plus stochastic differences amount to $\sim$ 2.5\% for these data for a cadence $\sim$~220\,s. B1 has a depth about 2.7\% (see Table \ref{Tab:AHSFit}), and it was not detected in the LCO light curve. For the late group, the Perkins data have the best quality followed by the FLWO KeplerCam data. 

As shown in Figs \ref{Fig: Optical_EG} and \ref{Fig: Optical_LG}, the overall agreement is quite good during the 26 hours, particularly for Dip B2 and B3. Dip B1 is isolated but it is the shallowest dip, and it is not well detected with smaller telescopes.

One main conclusion from this exercise is that in terms of observing a faint star like WD~1145+017, ``bigger is better". The big telescopes are good for capturing short timescale structure, whereas the small telescopes average over this structure. Fortunately, all telescopes measure the same overall structure and the depths of the dips, which we measure in this paper, are highly consistent among the different datasets.

\vskip 10mm
\noindent List of affiliations
\vskip 5mm
\noindent $^{1}$European Southern Observatory, Karl-Schwarzschild-Stra{\ss}e 2, D-85748 Garching, Germany\\
$^{2}$Department of Physics, and Kavli Institute for Astrophysics and Space Research, Massachusetts Institute of Technology, Cambridge, \\ MA 02139, USA \\
$^{3}$Institute of Astronomy, University of Cambridge, Madingley Road, Cambridge, CB3 0HA, UK \\
$^{4}$Harvard-Smithsonian Center for Astrophysics, 60 Garden Street, Cambridge, MA 02138 USA \\
$^{5}$Hereford Arizona Observatory, Hereford, AZ 85615, USA \\
$^{6}$School of Physics and Astronomy, Tel-Aviv University, Tel-Aviv 6997801, Israel\\
$^{7}$European Southern Observatory, Ave. Alonso de C\'{o}rdova 3107, Vitacura, Santiago, Chile\\
$^{8}$Visidyne, Inc., Santa Barbara, CA 93105, USA \\
$^{9}$Observatoire Astronomique de l'Universit'e de Geneve, 51 ch.~des Maillettes, 1290 Versoix, Switzerland \\
$^{10}$Research School of Astronomy and Astrophysics, Mount Stromlo Observatory, Australian National University, Weston, ACT 2611, Australia \\
$^{11}$Centre for Exoplanet Science, SUPA School of Physics \& Astronomy , University of St Andrews, North Haugh, ST ANDREWS KY16 9SS, UK\\
$^{12}$Department of Physics and Astronomy, George Mason University, Fairfax, Virginia, USA (GMU observations)\\
$^{13}$Institute for Astrophysical Research, Boston University, 725 Commonwealth Avenue, Room 506, Boston, MA 02215, USA \\
$^{14}$Space Telescope Science Institute, Baltimore, MD 21218, USA\\
$^{15}$Paul \& Jane Meyer Observatory, Coryell County, TX\\
$^{16}$Institut de Recherche sur les Exoplan\`etes (iREx) and D\'epartement de physique, Universit\'e de Montr\'eal, Montr\'eal, QC H3C 3J7, Canada \\
$^{17}$Eureka Scientific Inc., Oakland, CA, USA 94602\\
$^{18}$Instituto de Astrof\'isica, Facultad de F\'isica, Pontificia Universidad Cat\'{o}lica de Chile, Av.~Vicu\~na Mackenna 4860, 782-0436 Macul, \\ Santiago, Chile\\
$^{19}$Department of Physics and Astronomy, Brigham Young University, Provo, UT 84602, USA(BYU observations)\\
$^{20}$Department of Physics and Astronomy, University of California, Los Angeles, CA 90095-1562, USA \\
$^{21}$Raemor Vista Observatory, 7023 E. Alhambra Dr., Sierra Vista, AZ 85650, USA (JBO observations)\\
$^{22}$Temple College, Temple, TX, 76504\\
$^{23}$Instituto de Astrof\`{i}sica de Canarias, V\`{i}a L\`{a}ctea s/n, E-38205 La Laguna, Tenerife, Spain\\
$^{24}$Departamento de Astrof\`{i}sica, Universidad de La Laguna, Spain\\
$^{25}$Department of Physics and Astronomy, University of Delaware, Newark, DE 19716, USA\\
$^{26}$South African Astronomical Observatory, PO Box 9, Observatory, 7935, South Africa\\
$^{27}$Division of Geological and Planetary Sciences, California Institute of Technology, Pasadena, CA 91125, USA\\

\bsp	
\label{lastpage}
\end{document}